\documentclass[
 reprint,
 amsmath,amssymb,
 aps,
prb,
floatfix,
]{revtex4-2}

\usepackage{graphicx}
\usepackage{dcolumn}
\usepackage{bm}
\usepackage{hyperref}

\usepackage[justification=raggedright]{caption}
\usepackage{subcaption}

\usepackage{xcolor}
\usepackage{braket}
\usepackage[english]{babel}
\newcommand{\human}[1]{{\textsc{#1}}}
\begin{document}
\title{Scrambling and Many-Body Localization in the XXZ-Chain}
\author{Niklas B\"olter}
\author{Stefan Kehrein}
 \email{stefan.kehrein@theorie.physik.uni-goettingen.de}
\affiliation{
Institute for Theoretical Physics, Universität G\"ottingen, \\
Friedrich-Hund-Platz 1, 37077 G\"ottingen, Germany
}
\date{\today}
\begin{abstract}
The tripartite information is an observable-independent measure for scrambling and delocalization of information. Therefore one can expect that the tripartite information is a good observable-independent indicator for distinguishing between many-body localized and delocalized regimes, which we confirm for the XXZ-chain in a random field. Specifically, we find that the tripartite information signal spreads inside a light cone that only grows logarithmically in time in the many-body localized regime similar to the entanglement entropy. We also find that the tripartite information eventually reaches a plateau with an asymptotic value that is suppressed by strong disorder.
\end{abstract}
\maketitle
\section{Introduction}
Because of breakthroughs in the manipulation and observation of quantum systems well isolated from any environment - e.g. in cold atom experiments,  - the question of thermalization of a closed quantum system has become experimentally accessible \citep{blattQuantumSimulationsTrapped2012,blochManybodyPhysicsUltracold2008,dohertyNitrogenvacancyColourCentre2013,duanColloquiumQuantumNetworks2010,dumitrescuKosterlitzThoulessScalingManybody2019,kellyStatePreservationRepetitive2015,leibfriedQuantumDynamicsSingle2003,schirhaglNitrogenVacancyCentersDiamond2014,langenUltracoldAtomsOut2015,abaninColloquiumManybodyLocalization2019}.

Closed quantum systems undergo unitary time evolution, which implies that no information about the initial state will be lost and initially pure states of the entire system will stay pure. Recovering that information at a later time might require global measurements, however.

In these systems, the question of thermalization instead becomes: can the system serve as a bath for its subsystems? This is the case if small subsystems can be described by a thermal ensemble after long times and in the thermodynamic limit, which yields a definition of thermalization appropriate for closed quantum systems. Of particular interest is the thermalization of eigenstates of the Hamiltonian: the Eigenstate Thermalization Hypothesis (ETH)  predicts for a wide range of systems that the eigenstates of the Hamiltonian can serve as baths for their subsystems \citep{dalessioQuantumChaosEigenstate2016}.

In a similar vein, the unitary time evolution (which is in principle reversible by evolving backward in time) cannot destroy any information encoded in the system, since this would constitute an irreversible process. Instead, this information must still be stored in the system somehow, but hidden from local subsystems, since they are described by a thermal ensemble with a very small number of thermodynamic potentials insufficient to encode much information \citep{kaufmanQuantumThermalizationEntanglement2016}. Anyone able to measure arbitrary global observables on the other hand can, in theory, restore the information.

This motivates the concept of \emph{scrambling} in which initially local information delocalizes over the entire system. One way to quantify this are \emph{out-of-time-order correlation functions} (OTOCs) \citep{maldacenaBoundChaos2016}, which depend on a choice of observable, as well as the observable-independent \emph{tripartite information} $I_3$ \citep{hosurChaosQuantumChannels2016} that we will use in this article.

In the context of information spreading and thermalization, systems of interest are those that show a \emph{many-body localization} (MBL) transition \citep{oganesyanLocalizationInteractingFermions2007,gornyiInteractingElectronsDisordered2005,baskoMetalInsulatorTransition2006}, which have also been under investigation experimentally \citep{grossQuantumSimulationsUltracold2017,schreiberObservationManybodyLocalization2015,smithManybodyLocalizationQuantum2016,roushanSpectroscopicSignaturesLocalization2017,xuEmulatingManyBodyLocalization2018,rispoliQuantumCriticalBehaviour2019}. If a large amount of disorder is added to the system, all eigenstates become strongly localized, which implies that if initialized in such a state, the information does not spread across the system and it fails to thermalize (violates ETH). \citep{aletManybodyLocalizationIntroduction2018,abaninColloquiumManybodyLocalization2019,nandkishoreManyBodyLocalization2015,altmanUniversalDynamicsRenormalization2015}

Since the MBL phase is caused by strong disorder, adding additional small generic perturbations will not thermalize it, which makes this kind of ETH violation robust in comparison to \human{Bethe}-ansatz integrable systems which quickly become chaotic under perturbation, at least when diagnosed using level statistics \citep{znidaricWeakIntegrabilityBreaking2020}.

The robust localization of information in MBL systems could be useful in storing information in a quantum computer, which makes this an exciting topic not just for applying quantum information theory to MBL systems, but in the future using MBL systems in a quantum computer to in turn apply quantum information theory to new problems \citep{aletManybodyLocalizationIntroduction2018,bauerAreaLawsManybody2013}.

In this paper, we use the tripartite information to study the many-body localization transition. Our model system is the XXZ-chain in a random field tuned either to the non-interacting limit of the XX chain or the interacting Heisenberg chain. Using tools from quantum information theory we can calculate how information initially localized in a small part $\mathcal{A}$ of the system spreads into a distant small part $\mathcal{C}$ under time evolution. Changing the distance between $\mathcal{A}$ and $\mathcal{C}$ allows us to understand the spread of information in the system. 

Previous works established that in the MBL phase the half-chain entanglement entropy after a quantum quench spreads inside a light cone that is growing logarithmically with time, which distinguishes it from the \human{Anderson} localization in non-interacting systems without information spreading \citep{znidaricManybodyLocalizationHeisenberg2008,bardarsonUnboundedGrowthEntanglement2012,serbynUniversalSlowGrowth2013,andraschkoPurificationManyBodyLocalization2014,hopjanScalingPropertiesSpatial2021,huangExtensiveEntropyUnitary2021}.

The tripartite information was also used in \citet{maccormackOperatorEntanglementGrowth2021} to analyze, deep in the MBL phase, both an effective l-bit Hamiltonian as well as the disordered Heisenberg chain, where the logarithmically spreading light cone was already found. In our paper we also study the dependence of the delocalization on the disorder strength in both the interacting model with an MBL transition and the non-interactiong model that is Anderson localized and provide data for larger system sizes for the Heisenberg chain.

The tripartite information complements other quantities like the entanglement entropy, which is similarly observable-independent but does not directly measure the delocalization of information with spatial resolution. The converse is true for the OTOCs, which are not observable-independent. Other applications of the tripartite information also include the scrambling of product states \citep{iyodaScramblingQuantumInformation2018}, with suggestions on how this can be measured experimentally \cite{sunQuantumInformationScrambling2021}. Different from OTOCs and entanglement entropy the tripartite information also allows a quantitative comparison with a completely random system as described by (\human{Haar}) random time evolution as detailed in the next section \cite{hosurChaosQuantumChannels2016,schnaackTripartiteInformationScrambling2019}.

We find that the tripartite information $I_3$ can also be used to distinguish \human{Anderson} localization from both MBL and thermalization and that it also spreads inside a logarithmically growing light cone in the MBL phase. In addition, we see that the tripartite information reaches a plateau value eventually if the information is given enough time to explore the finite system, and this plateau value is also suppressed by disorder, so both the speed and the amplitude of the information transfer is inhibited in the MBL phase.

It should be noted that there is an ongoing discussion about the existence and nature of the MBL transition \citep{suntajsQuantumChaosChallenges2020,sierantCanWeObserve2021,abaninDistinguishingLocalizationChaos2021}, but since our goal is to do observable-independent studies, we are limited to smaller system sizes and cannot contribute to this discussion.

 The outline of our paper is as follows: In Sec.\,\ref{sec:scrambling} the tripartite information and scrambling are introduced, while in Sec.\,\ref{sec:MBL} we will quickly review many-body localization and get a first impression of the tripartite information in a disordered system. We will look at the non-interacting XX-chain in Sec.\,\ref{sec:XX} as a point of comparison, and then study the scrambling in the many-body localized phase in more detail in Sec.\,\ref{sec:XXZ} for the XXZ-chain in a random field. We conclude in Sec.\,\ref{sec:discussion}.

\section{Tripartite Information and Scrambling}
\label{sec:scrambling}
The delocalization of quantum information or \emph{scrambling} is a mechanism that describes how information that is initially localized becomes inaccessible for any local measurement. This allows even a closed quantum system, where the actual loss of information is prevented by unitary time evolution, to seemingly discard information.

While scrambling was initially discussed for black holes \citep{haydenBlackHolesMirrors2007,sekinoFastScramblers2008,lashkariFastScramblingConjecture2013,maldacenaBoundChaos2016}, we mostly follow \citet{hosurChaosQuantumChannels2016} to describe scrambling in unitary channels, in particular during unitary time evolution as described by the \human{von-Neumann} equation.

\begin{figure}[htb]
    \includegraphics[width=\columnwidth]{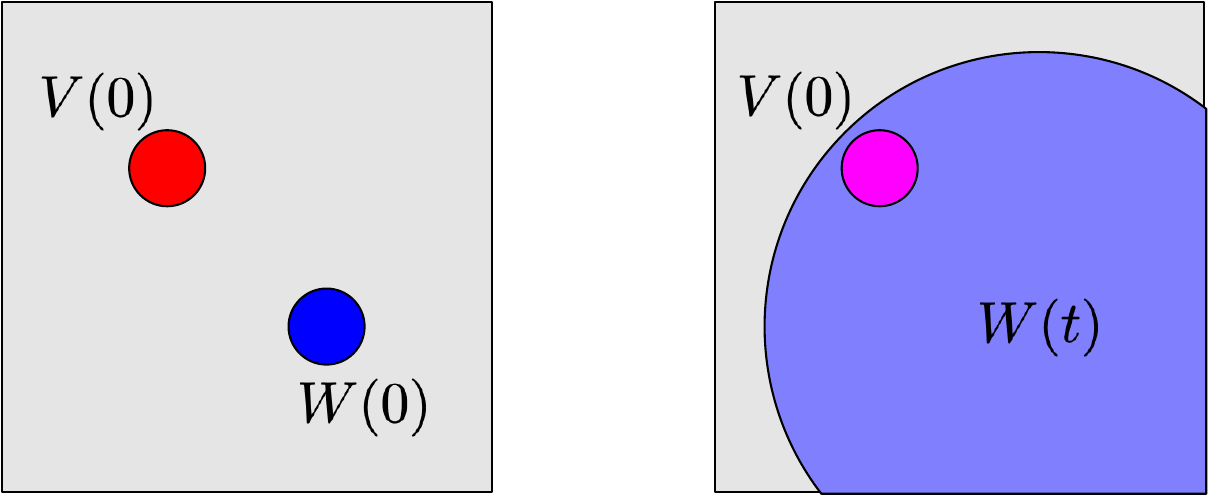}
    \caption{Spreading of initially local operator under time evolution.}
    \label{fig:Operator-Spreading}
\end{figure}

One well-known approach to scrambling are \emph{out-of-time-order correlation functions} (OTOCs). Two observables $\hat V$ and $\hat W$ that are initially localized in different parts of the system will spread over the system under scrambling dynamics as they evolve according to the non-linear Heisenberg equation (cf. Fig.\,\ref{fig:Operator-Spreading}). In mathematical terms, the commutator $\left[\hat W(t), \hat V(0)\right]$ which vanishes initially will become non-zero as $\hat W$ is evolved according to the Heisenberg equation. \citep{maldacenaBoundChaos2016}

This motivates one possible definition of quantum chaos which requires that the thermal expectation value $\left\langle \left[W(t), V(0)\right]^2 \right\rangle_\beta$ becomes of the order $2\langle VV\rangle_\beta \langle WW \rangle_\beta$.

The name OTOC comes from expanding the expression with the commutator and noticing that the interesting contributions come from out-of-time-order correlation functions like $\left\langle W(t)V(0)W(t)V(0) \right\rangle_\beta$. \citep{maldacenaBoundChaos2016}

The expression $\left\langle \left[W(t), V(0)\right]^2 \right\rangle_\beta$ can also, by replacing the commutator with a \human{Poisson} bracket in the semi-classical limit, be linked to a classical definition of chaos, where we see explicitly the dependence of the trajectory $q(t)$ on the initial conditions $q(0)$ and the \human{Lyapunov} exponent $\lambda$, which describes the exponential rate of separation in phase space in a chaotic classical system: \citep{maldacenaBoundChaos2016}
\begin{equation}
    \hbar^2 \left\{q(t), p\right\}^2 = \hbar^2 \left(\frac{\partial q(t)}{\partial q(0)}\right)^2 \propto \hbar^2 e^{2\lambda t}.
\end{equation}

The \emph{tripartite information} $I_3$ on the other hand was introduced in \citet{hosurChaosQuantumChannels2016} as an observable-independent measure of scrambling, which only depends on the partitioning of the Hilbert space of the quantum system into subsystems, and has also been linked to OTOCs.

\begin{figure}[htb]

\includegraphics[width=\columnwidth]{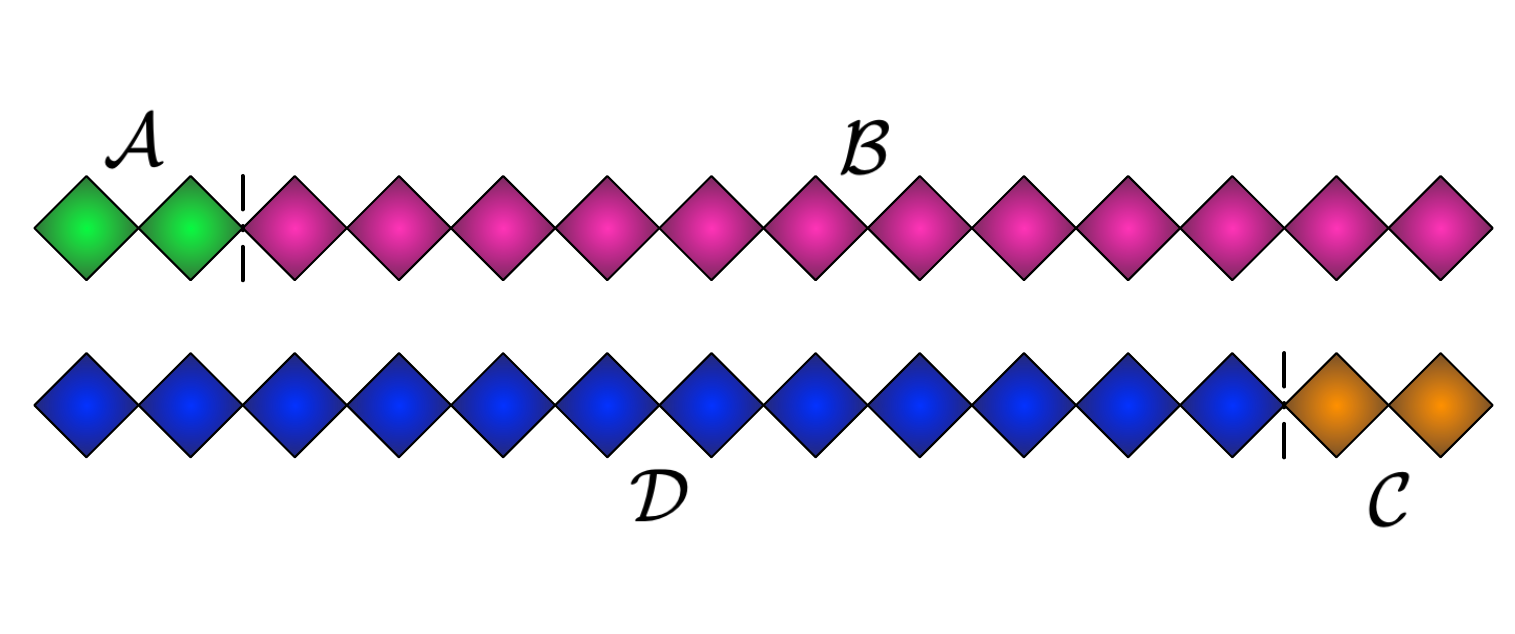}
    \caption{Subsystems at initial (top) and final (bottom) time. Each rhombus represents a spin-1/2 degree of freedom.}
    \label{fig:system-layout}
\end{figure}

The tripartite information requires a partitioning of the Hilbert space of the initial and final states of the system into two parts (bipartition):
\begin{align*}
    \rho(0) = \rho_\mathcal{AB} &\in \mathcal{H_A} \otimes \mathcal{H_B} \\
    \rho(t) = \rho_\mathcal{CD} &\in \mathcal{H_C} \otimes \mathcal{H_D}
\end{align*}
We will call the parts of the system described by these Hilbert spaces parts $\mathcal{A}$, $\mathcal{B}$, $\mathcal{C}$, $\mathcal{D}$ respectively.

For this purpose, we pick a small subset $\mathcal{A}$ of spins of our chain at the initial time $0$, and a small subset $\mathcal{C}$ of spins of our model at the final time $t$ as shown in Fig.\,\ref{fig:system-layout}, while all other spins are in the much larger subsystems $\mathcal{B}$ and $\mathcal{D}$. \citep{hosurChaosQuantumChannels2016, schnaackTripartiteInformationScrambling2019}

\begin{figure}[htb]
    \includegraphics[width=\columnwidth]{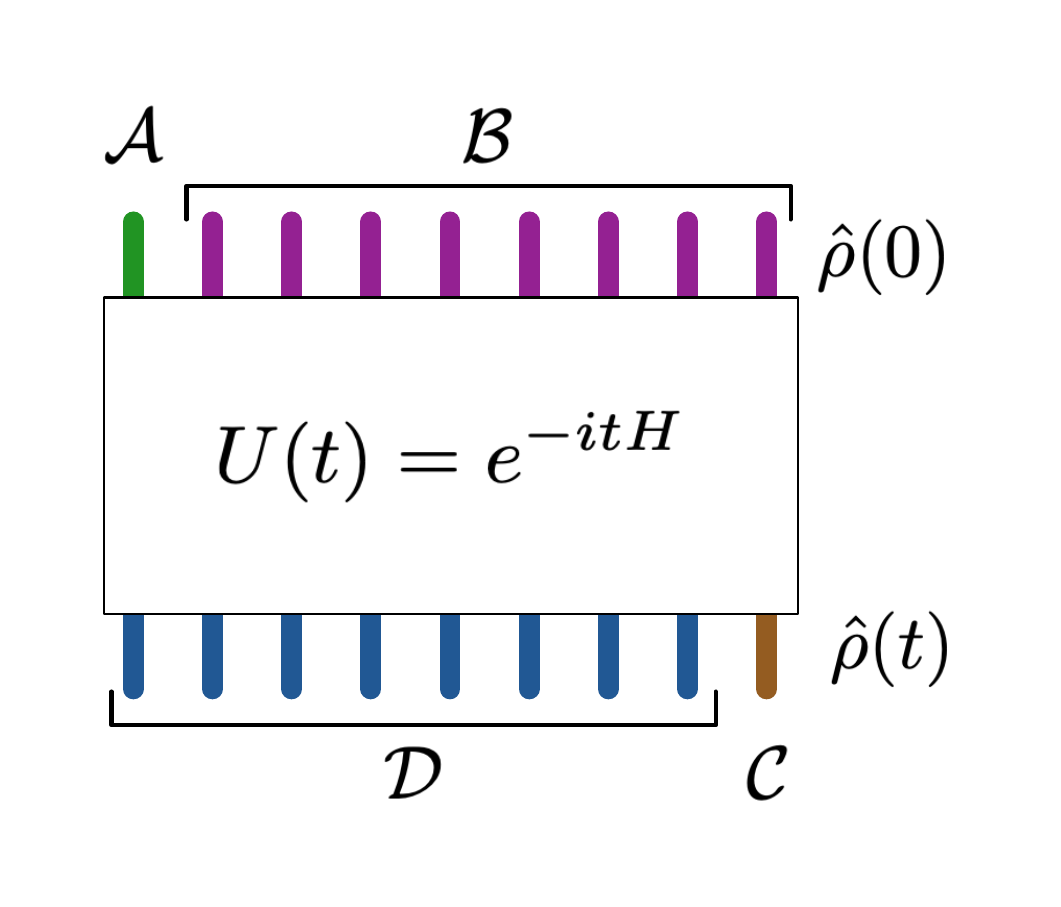}
    \caption{A quantum channel describing the time evolution of the quantum system and the subsystems used for calculating the tripartite information.}
    \label{fig:quantum-channel}
\end{figure}

In terms of quantum information theory, we map the time evolution operator to a unitary channel as shown in Fig.\,\ref{fig:quantum-channel}, where the spins at the initial time are now interpreted as input qubits described by the density matrix $\rho(0)$, and the spins at the final time as output qubits described by the time-evolved density matrix $\rho(t) = U(t) \rho(0) U^\dagger(t)$. Via channel-state duality we map this to a state in a doubled Hilbert space $\mathcal{H} \otimes \mathcal{H}$:
\citep{hosurChaosQuantumChannels2016}
\begin{equation}
    \ket{U(t)}_\mathcal{ABCD} = \sum_\nu \sqrt{\rho(0)} \ket{\nu}_\mathcal{AB} \otimes U(t) \ket{\nu}_\mathcal{CD}. \label{eq:tfd-state}
\end{equation}
Where $\left\{\ket{\nu}\right\}$ is an arbitrary orthonormal basis of $\mathcal{H}$, with the basis of product states like $\ket{\downarrow\downarrow},\ket{\downarrow\uparrow},\ket{\uparrow\downarrow},\ket{\uparrow\uparrow}$ (exemplary for the $L=2$ case) being a sensible choice to simplify calculating partial traces in the next step.

By construction, we have that tracing out the input or output system gives us the correct state:
\begin{align*}
    \mathrm{Tr}_\mathcal{CD} \ket{U(t)}\bra{U(t)} &= \rho_\mathcal{AB} = \rho(0), \\
    \mathrm{Tr}_\mathcal{AB} \ket{U(t)}\bra{U(t)} &= \rho_\mathcal{CD} = \rho(t).
\end{align*}
But more importantly, we can also trace over nonequal-time subsystems like $\mathcal{BD}$ to get a density matrix $\rho_\mathcal{AC}(t)$ that can, in general, be time-dependent even if the physical system is in thermal equilibrium - just because $\mathcal{AB}$ and $\mathcal{CD}$ are time-independent does not mean $\mathcal{AC}$ or $\mathcal{AD}$ are as well - and which describes how information is transported by the time evolution. We will use the \human{von Neumann} entanglement entropy $S_\chi = -\mathrm{Tr}\, \rho_\chi \log_2 \rho_\chi$ to calculate the mutual information between subsystems, here exemplary for correlations between A und C \citep{wildeQuantumInformationTheory2017}:
\begin{equation}
    I(\mathcal{A}:\mathcal{C}) = S_\mathcal{A} + S_\mathcal{C} - S_\mathcal{AC}.
\end{equation}
For our initial state, we will pick a density matrix mixing all states with zero-magnetization with equal probability (which corresponds to an infinite temperature ensemble in the largest symmetry block of our Hamiltonian). For this case, the initial entropies can be calculated analytically with some straight-forward combinatorics:
\begin{align*}
S(L,\ell)_{t=0} &= -\sum_{n=0}^\ell \begin{pmatrix}\ell\\n\end{pmatrix} p(L,l,n) \log_2 p(L,l,n), \\
p(L,l,n) &= \begin{pmatrix}L-\ell\\L/2-n\end{pmatrix} \begin{pmatrix}L \\L/2\end{pmatrix}^{-1}.
\end{align*}
Where $\ell$ is the size of the subsystem, as measured by the number of spins contained in it.
If we now want to calculate the tripartite information:
\begin{equation}
    I_3 = I(\mathcal{A:C}) + I(\mathcal{A:D}) - I(\mathcal{A:CD}),
    \label{eq:tripartite-information}
\end{equation}
it turns out we only need to calculate the entropies $S_\mathcal{AC}$ and $S_\mathcal{AD}$ of two nonequal-time subsystems numerically, since the other entropies are either time-independent or can be mapped to known ones via $S_\chi \equiv S_{\mathcal{ABCD}\setminus\chi}$. This is typically the case when the system starts in thermal equilibrium, as all equal-time subsystems are also in thermal equilibrium and time-independent (more entropies become time-dependent in e.g. a quench situation).

The tripartite information $I_3$ measures the delocalization of quantum information as it spreads from subsystem $\mathcal{A}$ over the entire system. This makes it a good approach to study MBL, in particular since it is observable-independent.

\citet{maccormackOperatorEntanglementGrowth2021} used the tripartite information to study the delocalization deep inside the MBL phase in an effective model with l-bits (cf. Eq.\,\ref{eq:lbits}), where a spreading inside a logarithmically growing light cone was found.  We note that the scrambling of information was also studied in \citet{banulsDynamicsQuantumInformation2017} by considering the mutual information between the first $\ell$ spins with the rest of the chain. The results however cannot be compared to ours as they use a specific initial state that singles out the first spin in the contrast to our ensemble of states in Eq.\,\ref{eq:tfd-state}.

The interpretation of the uniform incoherent superposition of all eigenstates of the Hamiltonian as the input state is not quite straightforward: we think of it as picking one eigenstate randomly at $t=0$ and then using the tripartite information to investigate if the particular choice can be revealed by an arbitrary local measurement at the later time $t$. This is also consistent with the fact that the tripartite information vanishes for pure input states, as in that case there is no choice about the initial state left.

For the infinite temperature ensemble, the tripartite information usually starts at zero and becomes negative under scrambling dynamics as either $I(\mathcal{A:C})$ or $I(\mathcal{A:D})$ become smaller, if there are no symmetry constraints and the full $2^L$ basis states are used in Eq.\,\ref{eq:tfd-state}. The negative value of the tripartite information is then normalized by dividing by the tripartite information $I_3^\textrm{Haar}$ of random time evolution (where the time evolution operator has been replaced by a random unitary sampled from the unitary group with the \human{Haar} measure \citep{schnaackTripartiteInformationScrambling2019,mezzadriHowGenerateRandom2007}). If the value after this normalization approaches $1$ the system is a scrambler for this particular partitioning of the Hilbert space. \citep{hosurChaosQuantumChannels2016,schnaackTripartiteInformationScrambling2019}

The tripartite information is bounded by the size of the smaller subsystem, $I_3 \geq -2\ell$, analogously to the entanglement entropy of a bipartite state \footnote{Consider that by construction $I_3 \geq -I(\mathcal{A:CD})$, and also $I(\mathcal{A:CD}) \leq 2 S_\mathcal{A} \leq 2 \log \mathrm{dim} \mathcal{H_A}$ by subadditivity and extremal properties of the \human{von-Neumann} entropy. }. Since we are not scaling $\ell$ with the system size $L$ but keep it fixed, the tripartite information cannot become extensive, in contrast to e.g. the half-chain entanglement entropy in the large $L$ limit. For $\ell = 2$ we find $I_3^\textrm{Haar} \approx -1.9$ for system sizes $8 \leq L \leq 14$.

In our case, the restriction to the zero-magnetization sector also introduces correlations in the initial state that shift the initial value of the tripartite information - we subtracted those so that $I_3$ again starts out at zero in our plots. Putting these two together means that we are plotting the following rescaled tripartite information:
\begin{equation}
    \tilde I_3(t) = \frac{I_3(t) - I_3(0)}{I_3^\textrm{Haar} - I_3(0)}.
\end{equation}

Other potential choices for the initial state are finite temperature ensembles, for which we however could not find a normalizing constant like the Haar averaged value $I_3^\textrm{Haar}$ to diagnose scrambling.

It is of note here that we are restricted to very small system sizes because we are storing the very large state of the system numerically exactly as well as doubling the system to two copies, which limits wide parameter studies with large numbers of samples to about $L = 12$ physical qubits \footnote{The system with 12 qubits uses about 1\,GiB of memory and 1 minute of CPU time on a current-generation CPU with fast BMI2 instructions to calculate one data point, with millions of data points needed for the plots in this article}.

\section{Many-Body Localization}
\begin{figure}[htb]
\includegraphics[width=0.8\columnwidth]{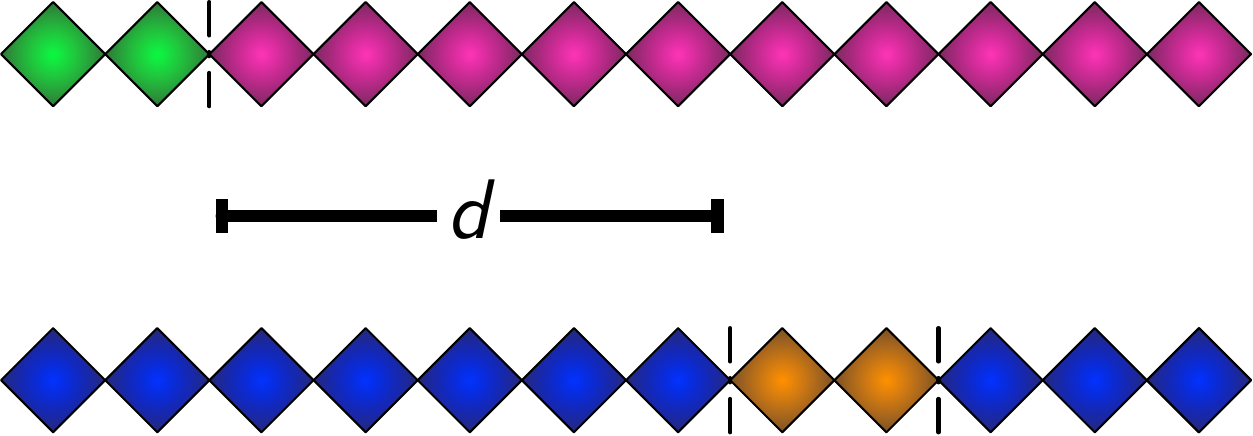}
    \caption{Illustration of the distance $d$ between subsystems $\mathcal{A}$ and $\mathcal{C}$.}
    \label{fig:subsystem-distance}
\end{figure}

\label{sec:MBL}
In this article, we will use the XXZ-Chain with open boundary conditions in a random and static magnetic field described by the following Hamiltonian: 
\begin{equation}
H = J \sum_{i=1}^{L-1}\left( S_i^x S_{i+1}^x + S_i^y S_{i+1}^y + \Delta S_i^z S_{i+1}^z \right) - \sum_{i=1}^{L} h_i S_i^z
\end{equation}
The magnetic field $h_i$ is taken randomly (independent, uniformly distributed) from $[-W,W]$. Since the random magnetic field introduces disorder we call $W/J$ the \emph{disorder strength}.

We will show some results for the non-interacting case of an XX-chain ($\Delta = 0$), but the bulk of our results was calculated for the isotropic ($\Delta = 1$) Heisenberg chain, which has a dynamical transition from thermalizing to the MBL regime at $W/J \approx 4$ for numerically accessible system-sizes \citep{beraManyBodyLocalizationCharacterized2015,maceMultifractalScalingsManyBody2019,aletManybodyLocalizationIntroduction2018,laflorencieChainBreakingKosterlitzThouless2020}. In the following, we will measure energy in units of $J$ and set $J = 1$.

Another class of closed systems that fail to thermalize in the above sense are \emph{integrable} models which have a large number of conserved quantities that restrict their evolution. Some of these quantities might be localized in a subsystem, which suppresses the scrambling of information as well \citep{schnaackTripartiteInformationScrambling2019}. These systems instead equilibrate to a so-called \emph{generalized Gibbs ensemble} \citep{rigolRelaxationCompletelyIntegrable2007,vidmarGeneralizedGibbsEnsemble2016}.

It turns out that the system in the MBL phase also has integrals of motion called l-bits that are localized around the position of physical qubit (p-bit) \citep{serbynUniversalSlowGrowth2013,husePhenomenologyFullyManybodylocalized2014,rosIntegralsMotionManybody2015}.
In terms of the l-bits $\tau_i$ the Hamiltonian can be rewritten as follows:
\begin{align}
H&=\sum_ih_i\tau^z_i + \sum_{i,j}J_{ij}\tau^z_i\tau^z_j \label{eq:lbits} \\ &+ \sum_{n=1}^{\infty} \sum_{i,j,\{k\}}K^{(n)}_{i\{k\} j}\tau^z_i\tau^z_{k_1}...\tau^z_{k_n} \tau^z_j, \nonumber
\end{align}
where the interaction terms fall off exponentially with distance. If a description as in Eq.\,\ref{eq:lbits} is valid, we do not expect scrambling as the information stays localized.

To study the spread of information through the spin chain we will pick two small subsystems, subsystem $\mathcal{A}$ for the initial time and subsystem $\mathcal{C}$ for the final time. To exploit the observable-independent nature of the tripartite information we would like to have subsystems that span multiple spins since they admit a larger number of independent observables that we will be able to describe collectively. So the size of the smaller subsystems $\ell$ should be large. On the other hand, we want to study the spread of information through the system, so the subsystem separation $d$ should also be varied over a wide range. Also, we are interested in \emph{local} information, which on its face requires that $\ell \ll L$.

It would have been desirable to put the two subsystems $\mathcal{A}$ and $\mathcal{C}$ far away from the ends of the chain to get rid of boundary effects. However, this is not possible due to the numerically reachable system sizes. We instead put the subsystem $\mathcal{A}$ right at an open boundary, which minimizes potential problems both with the reachable distances $d$ between the subsystems and reflections at the boundary that arrive at different times and complicate the analysis. We do not expect large boundary effects though, as they were not playing a large role in \citet{schnaackTripartiteInformationScrambling2019}, and we will see further below from Fig.\,\ref{fig:a3-asymptotic-change2} \& \ref{fig:a3-arrival-time-over-d} that they are rather weak as far as subsystem $\mathcal{C}$ is concerned. Boundary effects also typically only affect a few eigenstates significantly and should not play a large role in the $\beta \to 0$ ensemble.

So we will fix the small subsystem $\mathcal{A}$ at the left boundary and move the small subsystem $\mathcal{C}$ along the chain, with the distance $d$ between them (cf. Fig.\,\ref{fig:subsystem-distance}). This allows the distance $d$ to grow up to $L - 2 \ell$. To allow a wide range of subsystem separation $d$ while still retaining subsystems with multiple spins we fix $\ell = 2$ in the main part of this article. Some additional results for $\ell = 1,3$ that support our results in the main part can be found in the App.\,\ref{sec:l1-l3-results}.

It is of note that larger subsystem sizes also lead to much less variance in the tripartite information between different disorder realizations, which allows us to use a smaller sample size as would have been necessary for single-spin subsystems ($\ell = 1$).

\begin{figure}[htb]
    \includegraphics[width=\columnwidth]{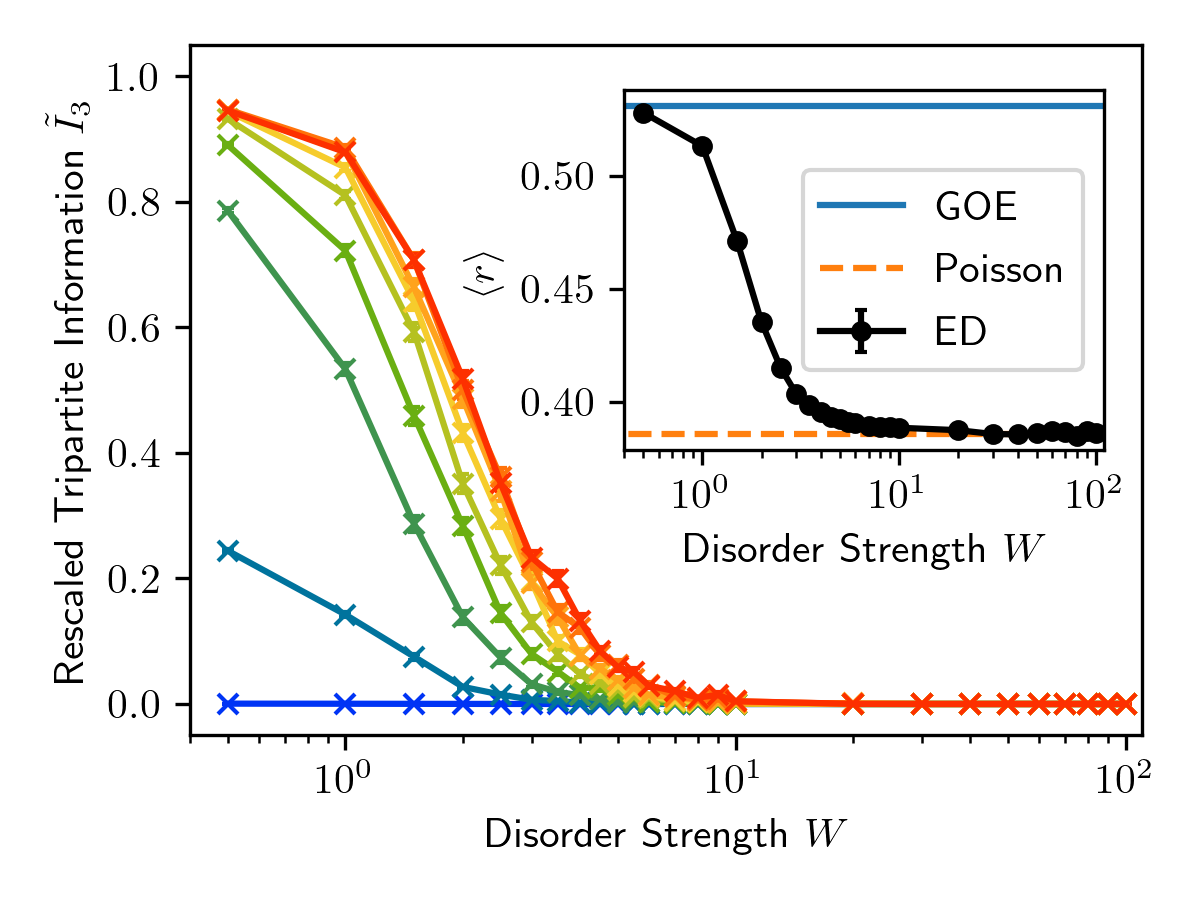}
    \caption{Comparison of the tripartite information between subsystems of size $\ell = 2$ at the opposite ends (cf. Fig.\,\ref{fig:system-layout}) of the isotropic Heisenberg chain ($\Delta = 1, L = 12$). The tripartite information is shown for different times $t = 4,8,16,32,64,128,256,512,1024$ from bottom (blue) to top (red). The inset shows the mean consecutive level spacing ratio $\langle r \rangle$ of the system's energy spectrum, which has been extensively studied in previous works, and tracks the late-time behaviour of tripartite information.}
    \label{fig:comparison-MBL-a3}
\end{figure}

\begin{figure}[htb]
    \includegraphics[width=\columnwidth]{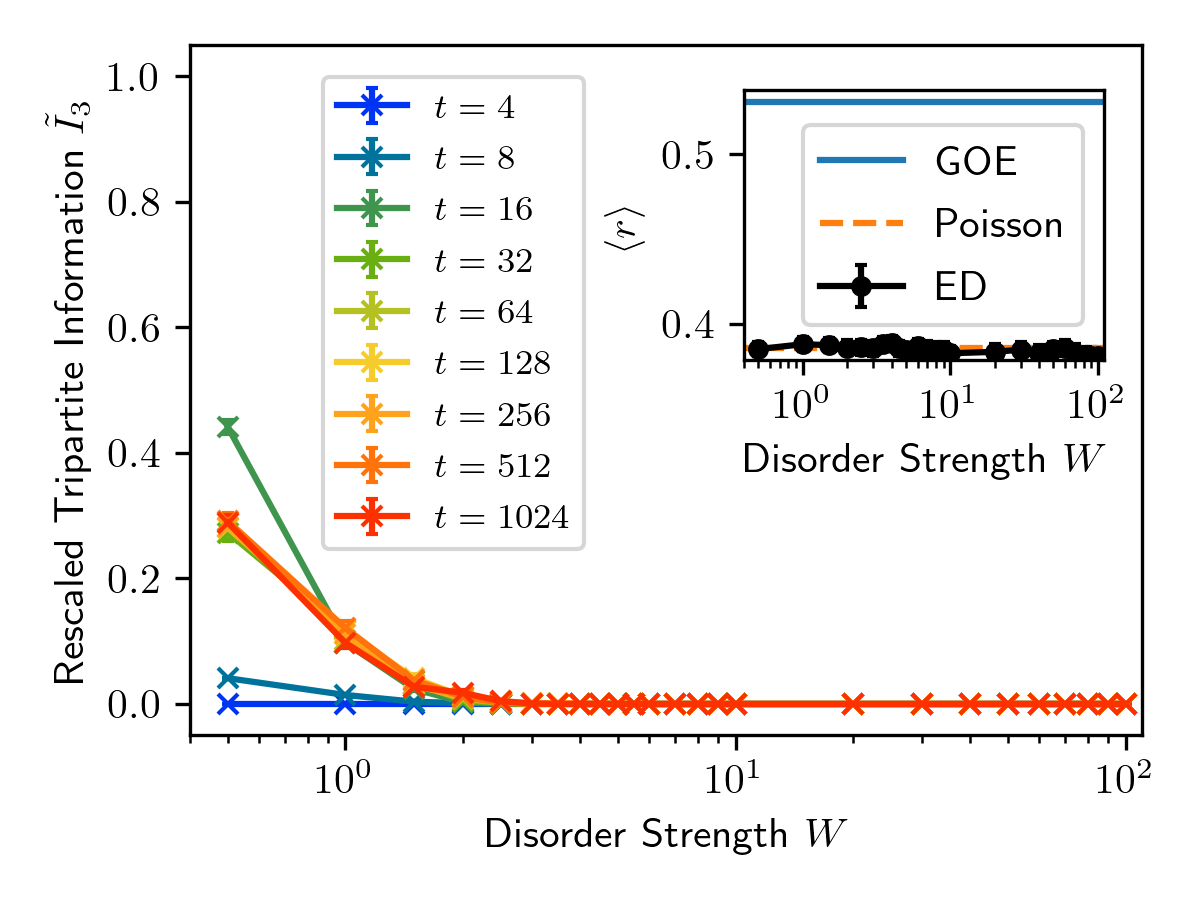}
    \caption{Comparison of the tripartite information in the non-interacting XX chain ($\Delta = 0, L = 12$) between subsystems of size $\ell = 2$ at the opposite ends  (cf. Fig.\,\ref{fig:system-layout}). The value of $\tilde I_3$ is only above 0.4 for $t = 16$ at low disorder. The inset shows the mean consecutive level spacing ratio $\langle r \rangle$ of the system's energy spectrum.}
    \label{fig:comparison-XX-a3}
\end{figure}

To get some first impressions of the tripartite information in disordered systems we calculated the mean consecutive level spacing ratio \citep{oganesyanLocalizationInteractingFermions2007,aletManybodyLocalizationIntroduction2018} and the tripartite information between subsystems at the ends of the chain at different times, both for the isotropic Heisenberg chain with MBL (Fig.\,\ref{fig:comparison-MBL-a3}) and the non-interacting system with Anderson localization (Fig.\,\ref{fig:comparison-XX-a3}).

As already found in previous works \citep{aletManybodyLocalizationIntroduction2018}, albeit with numerical data in systems of limited size, for the transition to MBL the mean consecutive level spacing ratio $\langle r \rangle$ goes from level repulsion comparable to that of the \emph{Gaussian Orthogonal Ensemble} (GOE) from random matrix theory at low disorder to non-correlated eigenvalues described by \human{Poisson} statistics at high disorder (cf. Fig.\,\ref{fig:comparison-MBL-a3}). In terms of localization, we expect that the system is in the localized regime for large disorder strength, which does suppress the spreading of information and hence the tripartite information, which is what we observe. For low disorder strength, the information can spread through the system. As soon as the information has had time to explore the finite system the tripartite information will saturate and approach the \human{Haar} value $\tilde I_3 = 1$.

\begin{figure}[htb]
    \includegraphics[width=\columnwidth]{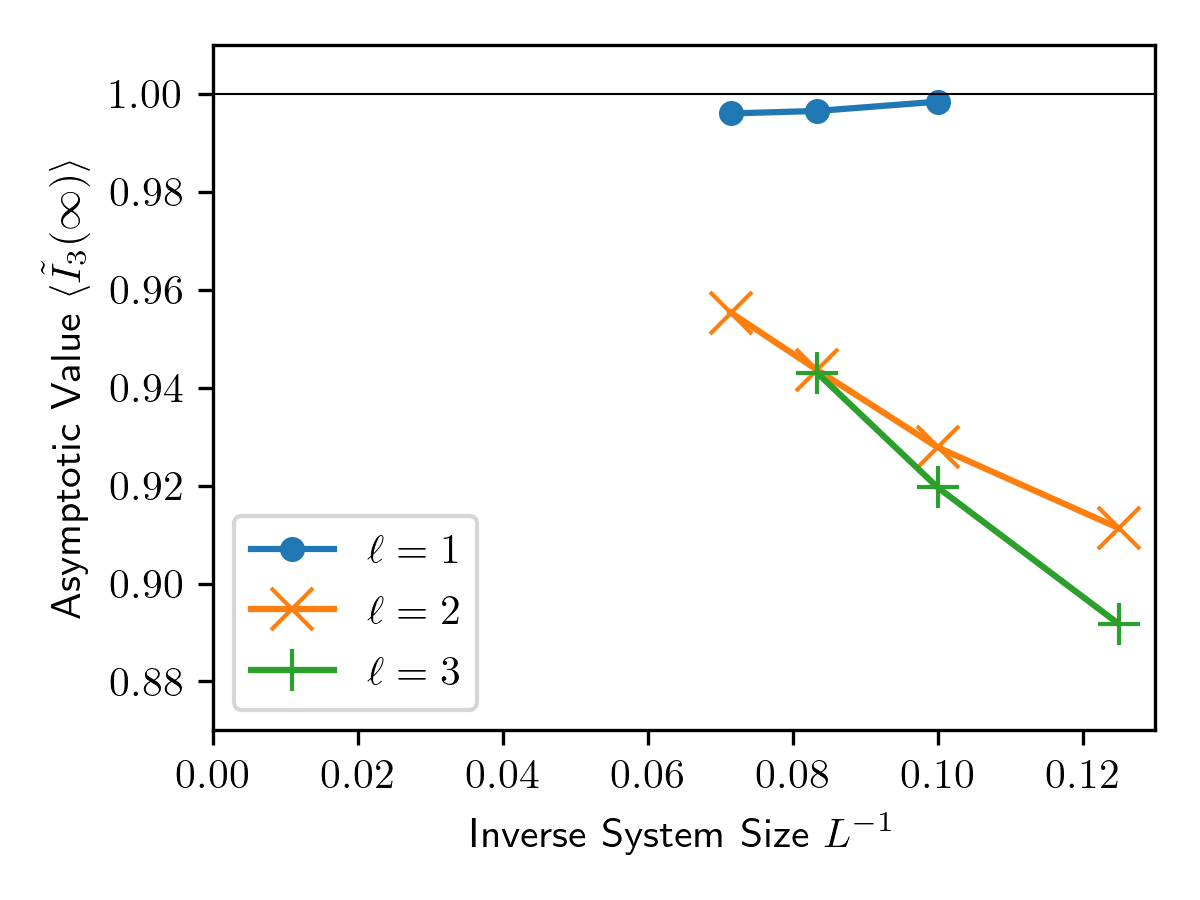}
    \caption{Highest asymptotic value of the rescaled tripartite information reached at late times for different system sizes $L$. The highest value is reached for low disorder strength $W$ (cf. Fig.\,\ref{fig:comparison-MBL-a3}).}
    \label{fig:asymptotic-finite-size}
\end{figure}

Although in our results the rescaled tripartite information stays below the scrambling value of $1.0$, we note that as we increase the system size from $L = 6$ to $L = 14$, this deviation shrinks from 30\,\% to 8\,\%, (cf. Fig.\,\ref{fig:asymptotic-finite-size}) which is consistent with a finite size effect that would vanish in the thermodynamic limit $L \to \infty$.

For the non-interacting XX-chain (Fig.\,\ref{fig:comparison-XX-a3}), which is \human{Anderson} localized, we see neither scrambling – since $I_3$ stays well below the \human{Haar} value – nor level-repulsion in the mean consecutive level spacing ratio $\langle r \rangle$ at low disorder, which is expected since the system stays localized in the absence of interactions. This also matches previous results for non-interacting systems, where a one-dimensional chain of fermions did not exhibit scrambling in the non-interacting limit \citep{schnaackTripartiteInformationScrambling2019}.

\section{Non-interacting XX-chain}
\label{sec:XX}

\begin{figure}[htb]
    \includegraphics[width=\columnwidth]{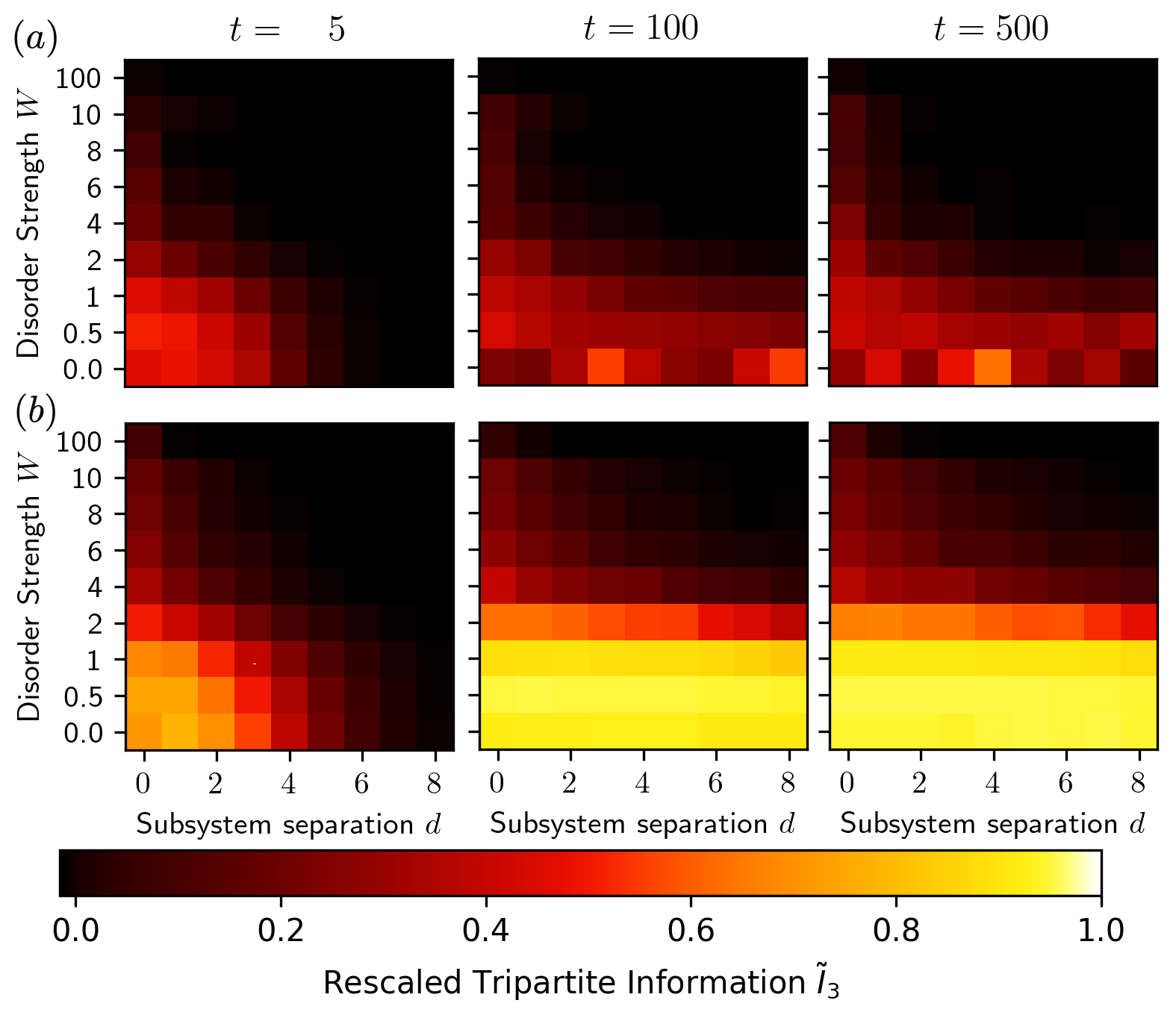}
    \caption{Spreading of information through (a) the XX chain ($\Delta = 0$) and (b) the isotropic Heisenberg chain ($\Delta = 1$) for different disorder strengths $W$ at early and late times of the ballistic spreading. Here the small subsystems $\mathcal{A}$ and $\mathcal{C}$ are of size $\ell = 2$ and the system size is $L = 12$. Full video available in the supplemental materials \citep{NoteSupplemental1,NoteSupplemental2}.}
    \label{fig:lightcone-movie-a3-XX-screencapture}
\end{figure}

\begin{figure}[htb]
    \includegraphics[width=\columnwidth]{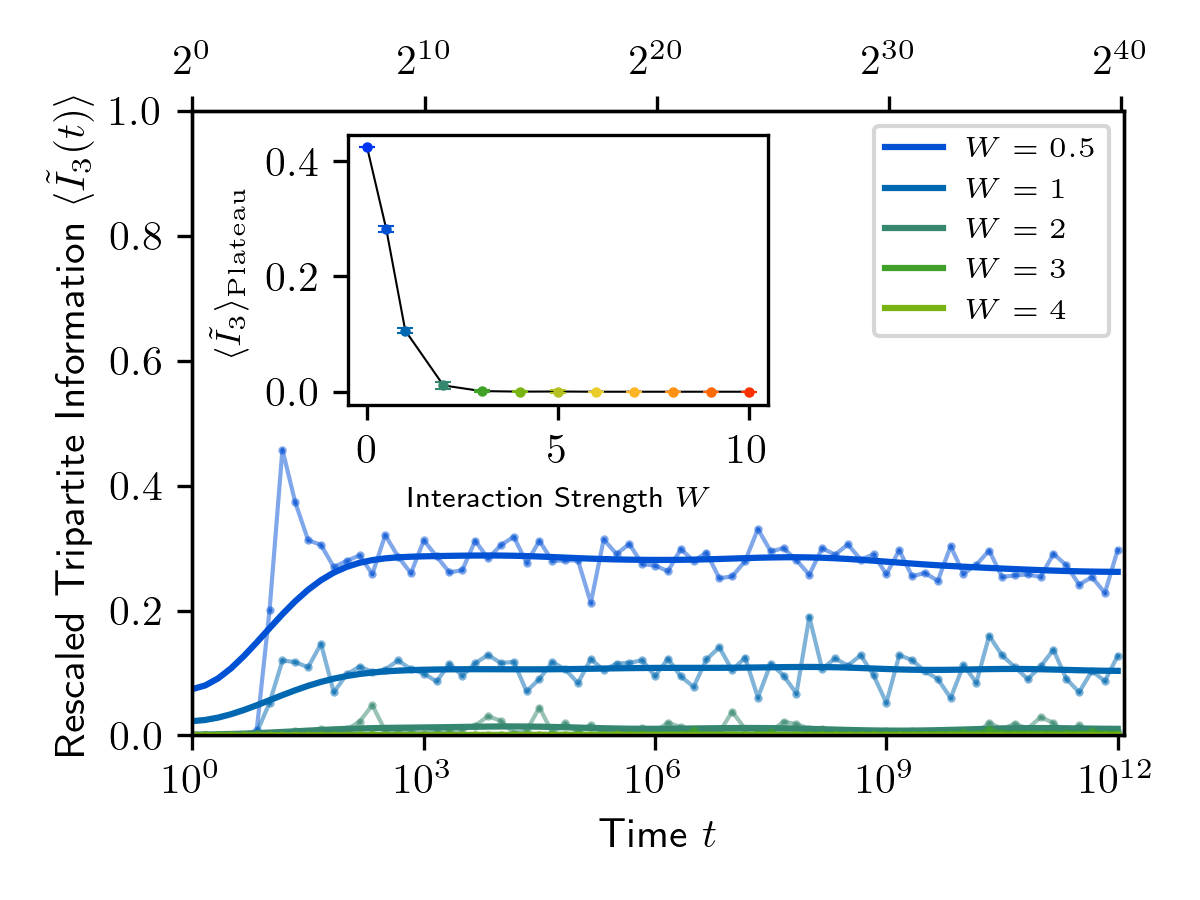}
    \caption{Spreading of information through the XX chain ($\Delta = 0, L = 12$) between two subsystems of size $\ell = 2$ at opposite ends. Because of the strong fluctuations a moving average (thick line) was added by convolution with a gaussian. The subset shows the height of the plateau, which was calculated by averaging data for intermediate times $10^3 \leq t \leq 10^{10}$.}
    
    \label{fig:a3-XX-logtime}
\end{figure}
The spreading of information through the non-interacting XX-chain is shown in Fig.\,\ref{fig:lightcone-movie-a3-XX-screencapture} (a) \& Fig.\,\ref{fig:a3-XX-logtime}. Similar to previous results in \citet{schnaackTripartiteInformationScrambling2019}, without interactions the tripartite information is fluctuating well below the \human{Haar} averaged value. In addition, we see that information transport is quickly suppressed by disorder, and no growth proportional to $\log(t)$ can be found in this system in contrast to the XXZ chain.

At later times than shown we observe unphysical behaviour in the tripartite information due to numerical errors, for that reason we did not consider times larger than $t = 10^{12}$ in our analysis.

\section{Scrambling in the XXZ-Chain}
\label{sec:XXZ}
Some qualitative overview is shown in Fig.\,\ref{fig:lightcone-movie-a3-XX-screencapture} (b), of which also a full video is available. For low disorder strength, the information is traveling through the system ballistically and becomes scrambled on a timescale linear in the system size \citep{kimBallisticSpreadingEntanglement2013}. In the MBL regime at higher disorder strength, the spreading is much slower and it will take a very long time until the signal hits the rightmost column in the video, which represents the correlations between the opposite ends of the system. We will analyse those correlations in more detail in the next paragraph.

\begin{figure}[htb]
    \includegraphics[width=\columnwidth]{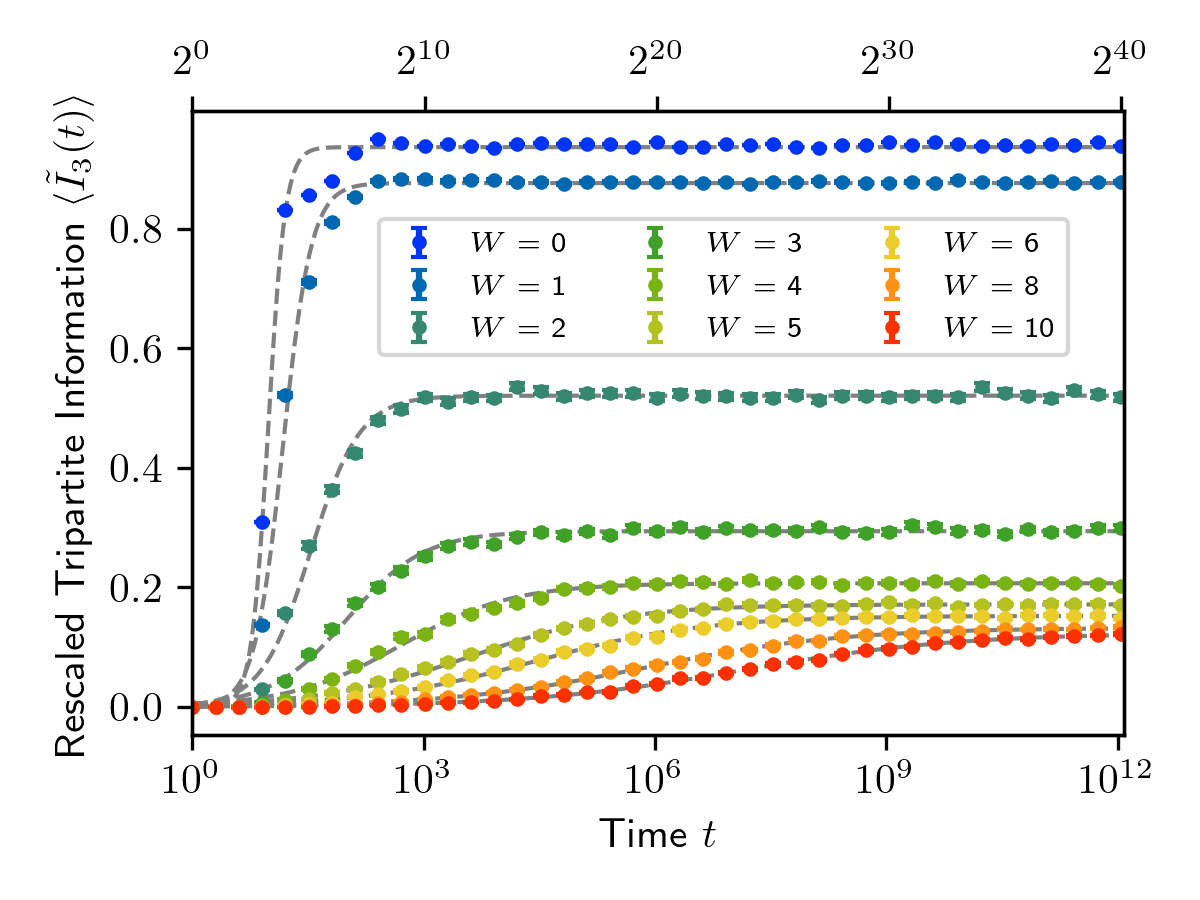}
    \caption{Spreading of information through the isotropic Heisenberg chain ($\Delta = 1, L = 12$) between two subsystems of size $\ell = 2$ at opposite ends. The dashed lines are fits against a logistic function, see main text.}
    \label{fig:a3-logtime}
\end{figure}

For some quantitative analysis, we calculated the tripartite information for subsystems at opposite ends of the chain and fitted them against a logistic function \footnote{Logistic function from \url{https://lmfit.github.io/lmfit-py/builtin_models.html\#stepmodel}} (cf. Fig.\,\ref{fig:a3-logtime}). We find that the tripartite information reaches a plateau after a time that becomes exponentially large deeper in the MBL phase in agreement with \citet{maccormackOperatorEntanglementGrowth2021}. The asymptotic value of $\tilde I_3$ at very late times (the height of the plateau) was estimated as well as the arrival time $T$ at which the change in the tripartite information hit half of that asymptotic value. In contrast to the non-interacting model from Fig.\,\ref{fig:a3-XX-logtime} we can see that the arrival time is pushed to exponentially large values for higher disorder in agreement with results for the slow rise of the entanglement entropy after a quantum quench \citep{znidaricManybodyLocalizationHeisenberg2008,bardarsonUnboundedGrowthEntanglement2012,serbynUniversalSlowGrowth2013,andraschkoPurificationManyBodyLocalization2014,hopjanScalingPropertiesSpatial2021,huangExtensiveEntropyUnitary2021}. Similar analyses were carried out for different values of the subsystem separation $d$. The plateau value and the timescale will be analysed in more detail below.

\begin{figure}[htb]
    \includegraphics[width=\columnwidth]{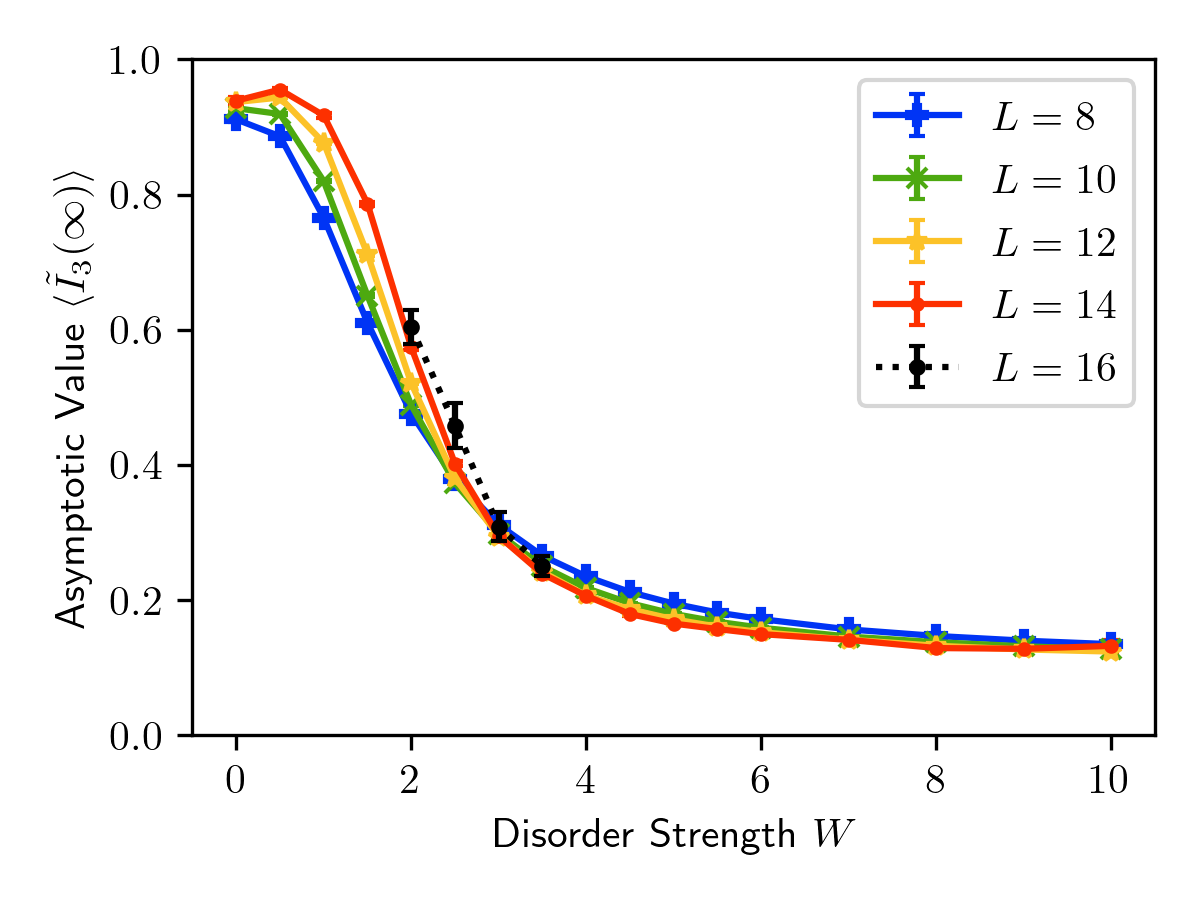}
    \caption{Asymptotic value at late times of the rescaled tripartite information $\tilde I_3(t)$ between two subsystems of size $\ell = 2$ at opposite ends of the isotropic Heisenberg chain ($\Delta = 1$). Here the system size is varied while $d$ is fixed at the maximum value $d = L - 2 \ell$. See Fig.\,\ref{fig:a3-asymptotic-change2} for results with varying subsystem separation.}
    \label{fig:a3-asymptotic-change}
\end{figure}

The asymptotic value of $\tilde I_3$ is shown in Fig.\,\ref{fig:a3-asymptotic-change}, where we see that the system at late time transitions from mostly delocalized at low disorder strength, where $\tilde I_3$ is close to the \human{Haar} scrambling value, to mostly localized at high disorder strength, where $\tilde I_3$ is far below the \human{Haar} scrambling value. The system size does not seem to have a large impact on the results, but the transition becomes sharper for larger systems. As we are dealing with finite systems, we do not expect a sharp transition to be visible between the thermalizing and the MBL phase, but instead a continuous transition to a regime with MBL characteristics \citep{morningstarAvalanchesManybodyResonances2021}. We can see however, that the curves cross each other at $W \approx 3$ in Fig.\,\ref{fig:a3-asymptotic-change}, which is of similar size compared to the literature value of the critical disorder strength $W_\mathrm{c} \approx 4$ \citep{beraManyBodyLocalizationCharacterized2015,maceMultifractalScalingsManyBody2019,aletManybodyLocalizationIntroduction2018,laflorencieChainBreakingKosterlitzThouless2020}. We note that there is an ongoing discussion about the nature of the transition, which we are unable to contribute to because our focus on the observable-independent tripartite information limits the accessible system sizes, nevertheless we also fitted our data to a Berezinskii-Kosterlitz-Thouless (BKT)-style curve in App.\,\ref{sec:bkt-scaling} following \citet{suntajsErgodicityBreakingTransition2020}.

\begin{figure}[htb]
    \includegraphics[width=\columnwidth]{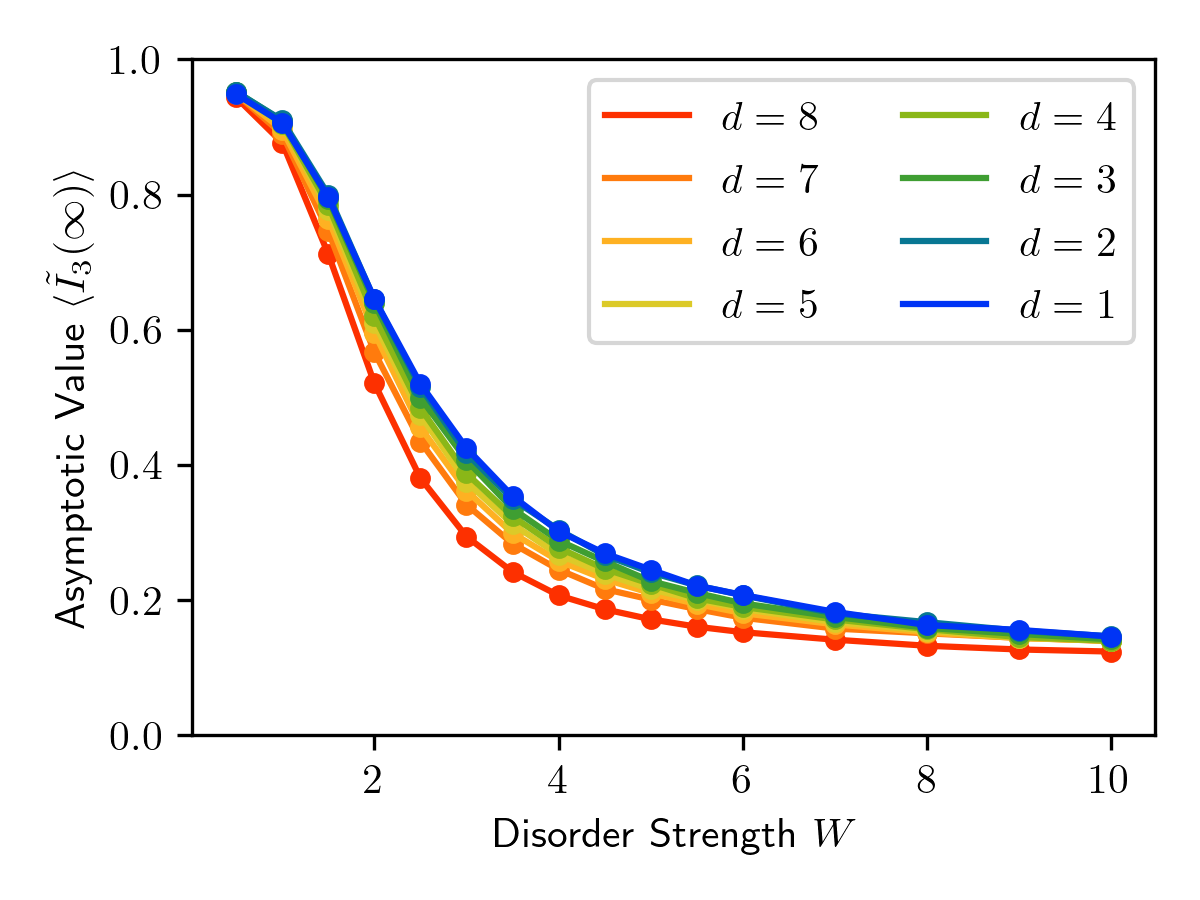}
    \caption{Asymptotic value at late times of the rescaled tripartite information $\tilde I_3(t)$ between two subsystems of size $\ell = 2$ of the isotropic Heisenberg chain ($\Delta = 1$). Here the subsystem separation is varied with $L = 12$ fixed, in contrast to the previous figure.}
    \label{fig:a3-asymptotic-change2}
\end{figure}

The asymptotic value of $\tilde I_3$ is also shown in Fig.\,\ref{fig:a3-asymptotic-change2}, this time for different values of the subsystem separation $d$. We expect that the tripartite information is suppressed if the distance $d$ becomes comparable to the localization length of the l-bits in Eq.\,\ref{eq:lbits}, and indeed $\tilde I_3$ shrinks as the distance is increased.

\begin{figure}[htb]
    \includegraphics[width=\columnwidth]{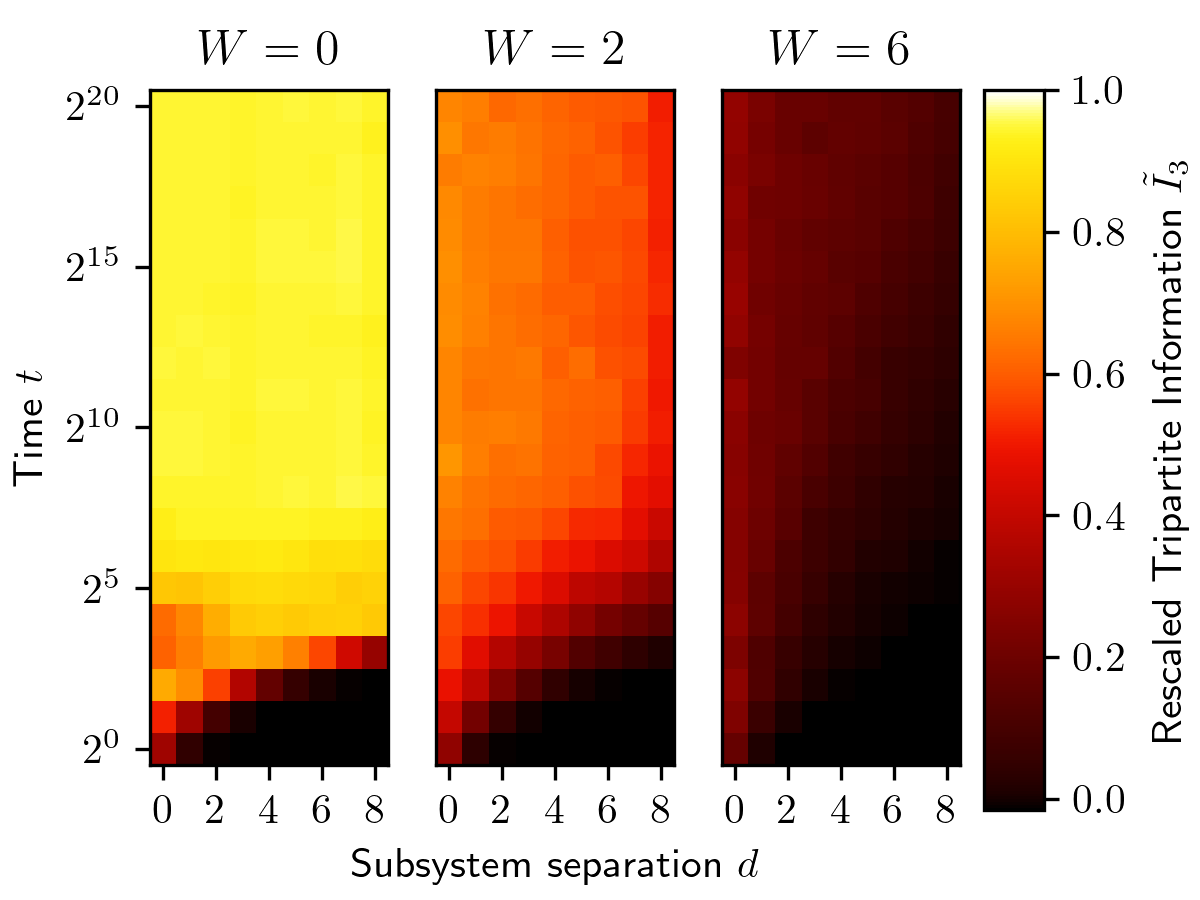}
    \caption{Dynamics of the tripartite information $I_3$ in the space-time plane with subsystems of size $\ell = 2 $ in the isotropic Heisenberg chain ($\Delta = 1, \ell = 12$). In the strongly disordered system ($W = 6$), the spreading of information takes an exponentially long time.}
    \label{fig:XX_W0_2_6}
\end{figure}

The light-cone is shown for three characteristic cases in Fig.\,\ref{fig:XX_W0_2_6}. In the strongly disordered system ($W = 6$), the spreading of information takes an exponentially long time.

\begin{figure}[htb]
    \includegraphics[width=\columnwidth]{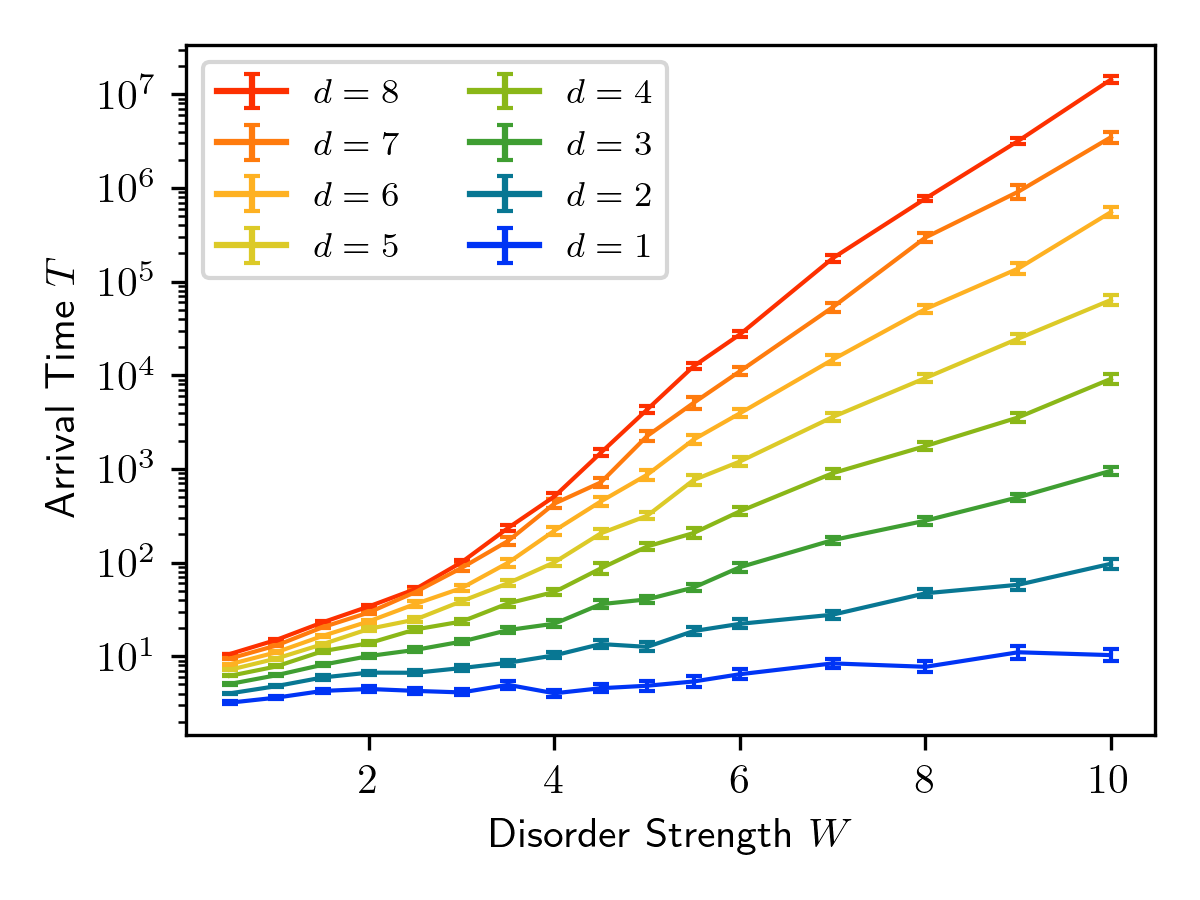}
    \caption{Arrival time of the information between two subsystems of size $\ell = 2$ of the isotropic Heisenberg chain ($\Delta = 1$) across a distance of $d$ spins. We again vary the subsystem separation $d$ for a fixed system size $L = 12$, a comparison of different system sizes can be found in Fig.\,\ref{fig:a3-arrival-time-over-d}.}
    \label{fig:a3-arrival-time}
\end{figure}

\begin{figure}[htb]
    \includegraphics[width=\columnwidth]{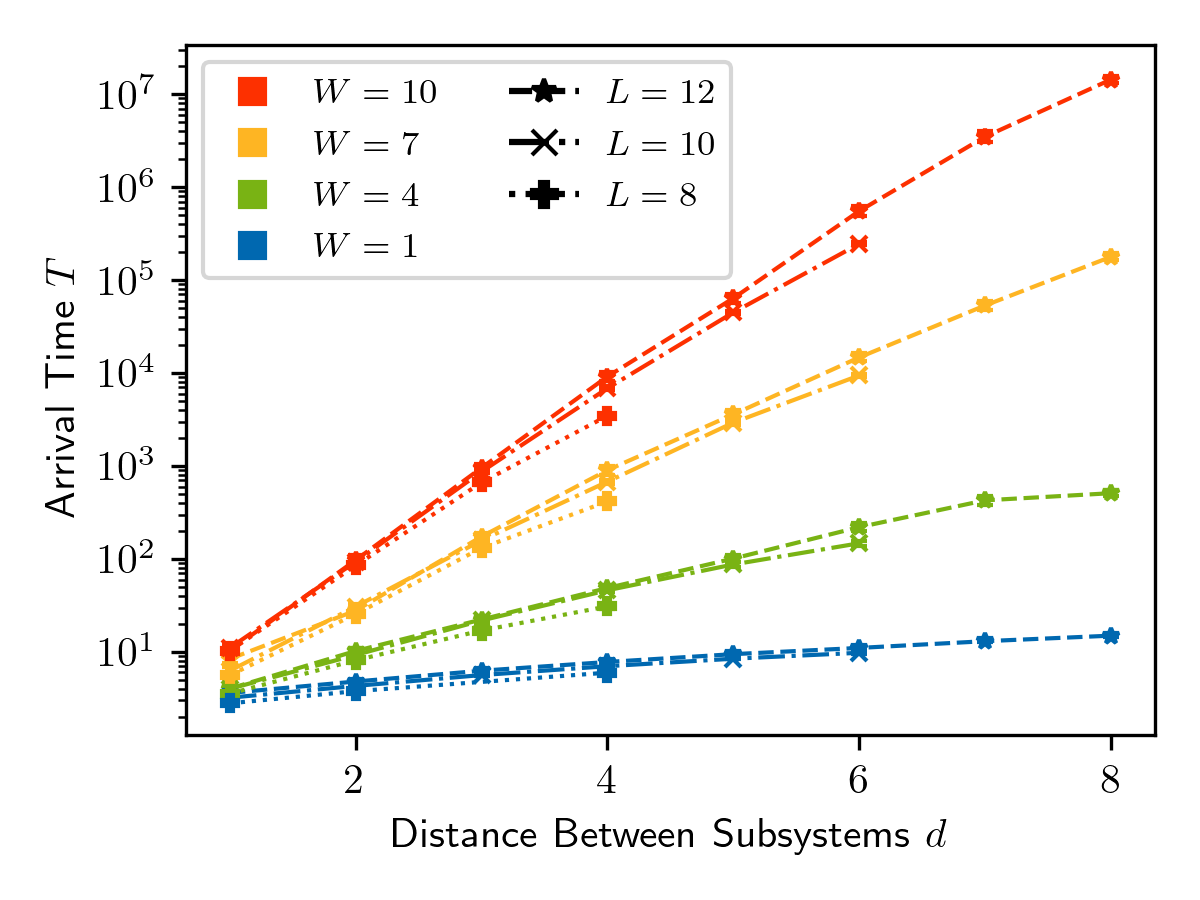}
    \caption{Arrival time of the information between two subsystems of size $\ell = 2$ of the isotropic Heisenberg chain ($\Delta = 1$) across a distance of $d$ spins.}
    \label{fig:a3-arrival-time-over-d}
\end{figure}

The arrival time $T$ on the other hand is shown in Fig.\,\ref{fig:a3-arrival-time} \& \,\ref{fig:a3-arrival-time-over-d},  where we see that the signal becomes exponentially slow as the distance or the disorder is increased. Since the arrival time  over distance data follows a straight line in the semi-log plot, we can deduce that the tripartite information spreads inside a light cone that grows logarithmically in time for large disorder. For the maximal distance between subsystems where $\mathcal{A}$ and $\mathcal{C}$ are right at the ends of the system, the arrival time is suppressed because of boundary effects, where the signal is reflected at the open boundary.
This causes the $d = 7$ and $d = 8$ curves in  Fig.\,\ref{fig:a3-arrival-time} to be slightly closer together than the others.

The boundary effect can also be seen quite clearly in Fig.\,\ref{fig:a3-arrival-time-over-d}, where the rightmost data points of the individual data set corresponding to subsystem $\mathcal{C}$ being at the boundary are slightly lower. As long as these boundary effects do not become important the arrival time does not depend strongly on the system size but only on the distance between the subsystems.

In summary, we can see that strong disorder suppresses both the amplitude of transmitted information as well as its speed.

\section{Discussion}
\label{sec:discussion}
We studied the spreading and delocalization of information in the random field XXZ model at the Heisenberg point $\Delta = 1$. Similar to previous results we see a clear change in behaviour as the information spreading is suppressed by strong disorder and the system becomes localized  \citep{devakulEarlyBreakdownAreaLaw2015,khemaniTwoUniversalityClasses2017,doggenManybodyLocalizationDelocalization2018,chandaTimeDynamicsMatrix2020,suntajsQuantumChaosChallenges2020,sierantThoulessTimeAnalysis2020,pandaCanWeStudy2020,chandaManybodyLocalizationTransition2020,suntajsErgodicityBreakingTransition2020,abaninDistinguishingLocalizationChaos2021}. In the MBL phase the information spreads slowly, contained inside a light cone that grows logarithmically with time, similar to results for the rise of the entanglement entropy after a quantum quench \citep{znidaricManybodyLocalizationHeisenberg2008,bardarsonUnboundedGrowthEntanglement2012,serbynUniversalSlowGrowth2013,andraschkoPurificationManyBodyLocalization2014,hopjanScalingPropertiesSpatial2021,huangExtensiveEntropyUnitary2021} and results for the tripartite information in the effective l-bit model \citep{maccormackOperatorEntanglementGrowth2021}. This logarithmic light cone is absent in the \human{Anderson} localized XX model.
We also found that the amplitude of the signal is suppressed since $\tilde I_3$ reaches a plateau value well below that of the delocalized system even after the signal has explored the finite system.

Because of the observable-independent nature of $I_3$ we can see the generic information transfer between subsystems that contain multiple spins without the difficulty of selecting a representative sample of local operators.

The tripartite information in our calculations is bounded by the fixed subsystem size and hence by construction no area or volume laws can be found, which distinguishes it from the half-chain entanglement entropy in the large $L$ limit \citep{aletManybodyLocalizationIntroduction2018}. We see instead quite clearly that for the non-interacting XX-chain in Fig.\,\ref{fig:a3-XX-logtime} the tripartite information vanishes as disorder causes \human{Anderson} localization.

The results for the asymptotic value of the tripartite information in the interacting XXZ-chain (cf. Fig.\,\ref{fig:a3-asymptotic-change}) are similar to \citet{khemaniCriticalPropertiesManyBody2017}, where the entanglement entropy of a single spin was used, and it was shown that it also stays well below the thermal value in the MBL phase.

The suppression of an asymptotic value in MBL systems was also found in \citet{raySignatureChaosDelocalization2018} (see also \citep{garcia-mataChaosSignaturesShort2018}) using OTOCs in a \human{Floquet} system, although a similar behaviour in the disordered Heisenberg chain was found to be a finite size effect \citep{chenOutoftimeorderCorrelationsManybody2017a}.

We observed exponential slowdown of the information transport in the MBL phase, similar results were also found in \citet{detomasiQuantumMutualInformation2017} for the mutual information between individual sites in the spinless fermionic disordered Hubbard chain, which showed logarithmically slow growth.

In this article we have used the \human{von-Neumann} entropy $S = -\mathrm{Tr} \rho \log_2 \rho$  throughout to calculate the mutual information instead of the related second \human{Rényi} entropy $S = -\log_2 \mathrm{Tr} \rho^2$, which has been linked to OTOCs \citep{hosurChaosQuantumChannels2016} and is also used to diagnose scrambling \citep{sunderhaufQuantumChaosBrownian2019}.  However, we also calculated the tripartite information $I_3^{(2)}$ corresponding to this entropy (data not shown), and found no qualitative difference in the behaviour.

The tripartite information can make powerful observable-independent statements about scrambling, but we think that it is difficult to access in an experiment, similar to the \human{von-Neumann} entanglement entropy. Experimental realizations might require a generalization of the initial state to finite temperature.

In summary, the tripartite information $I_3$ can make powerful, observable-independent and quantitative statements about scrambling. Specifically in this paper, we have found that the tripartite information allows us to study MBL, and also allows us to distinguish it from \human{Anderson} localization.

\section{Acknowledgements}
We acknowledge helpful discussions with Fabian Heidrich-Meisner and Alexander Osterkorn. This work was funded by the Deutsche Forschungsgemeinschaft (DFG, German Research Foundation) - 217133147/SFB 1073, project B03. The code to calculate partial traces was tested with the help of Oskar Schnaack and Sebastian Paeckel. Our numerical code uses the LAPACK \citep{andersonLAPACKUsersGuide1999} implementations from Intel MKL \footnote{\url{https://software.intel.com/oneapi/onemkl}} and AMD AOCL \footnote{\url{https://developer.amd.com/amd-aocl/}}.
\bibliography{MBL-Paper}

\begin{appendix}
\section{Results for subsystems of length 1 or 3}
\label{sec:l1-l3-results}
In addition to the behaviour of the tripartite information $I_3$ for a subsystem size $\ell = 2$ in the main part, we also studied smaller $(\ell = 1)$ and larger $(\ell = 3)$ subsystems.

\begin{figure}[htb]
    \subcaptionbox{$\ell = 1$\label{fig:a1-logtime}}{\includegraphics[width=0.95\columnwidth]{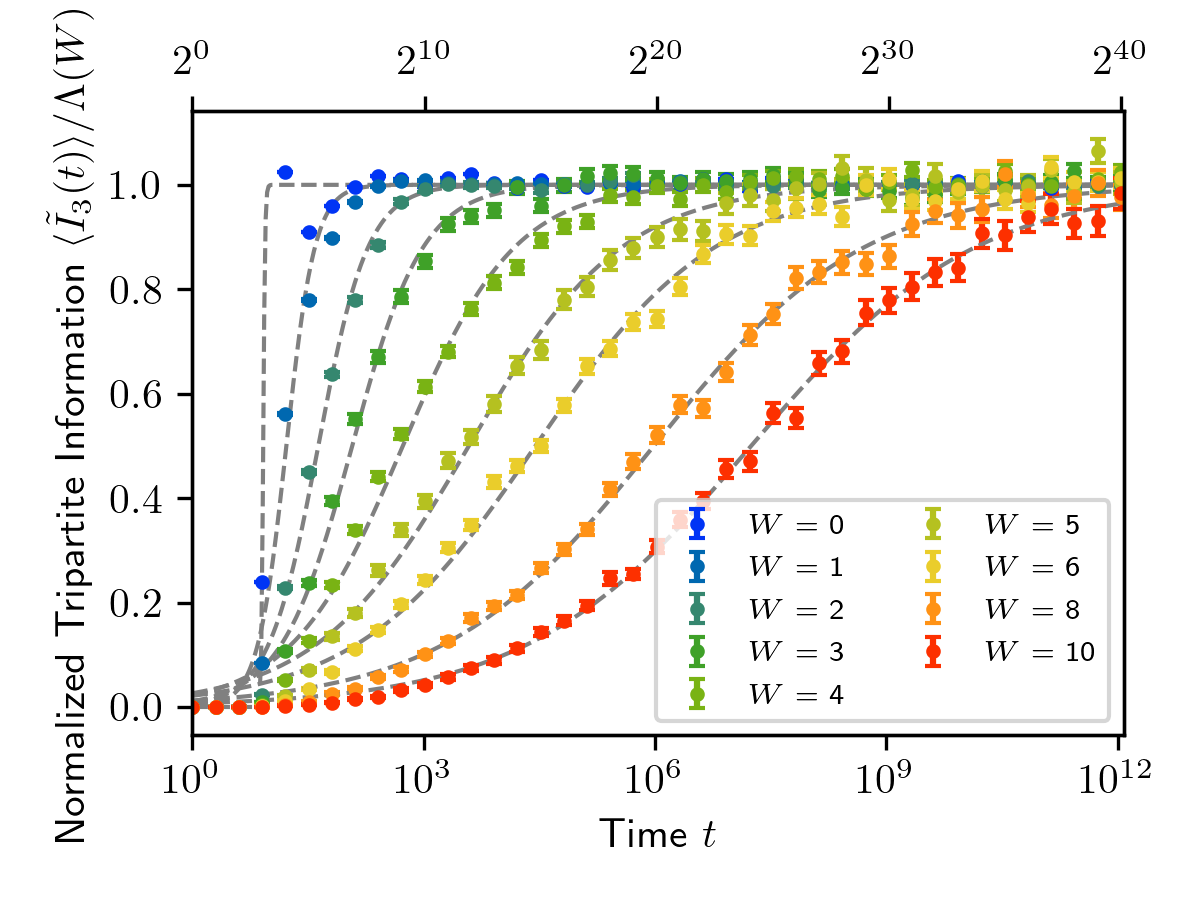}}
    \subcaptionbox{$\ell = 3$\label{fig:a7-logtime}}{\includegraphics[width=0.95\columnwidth]{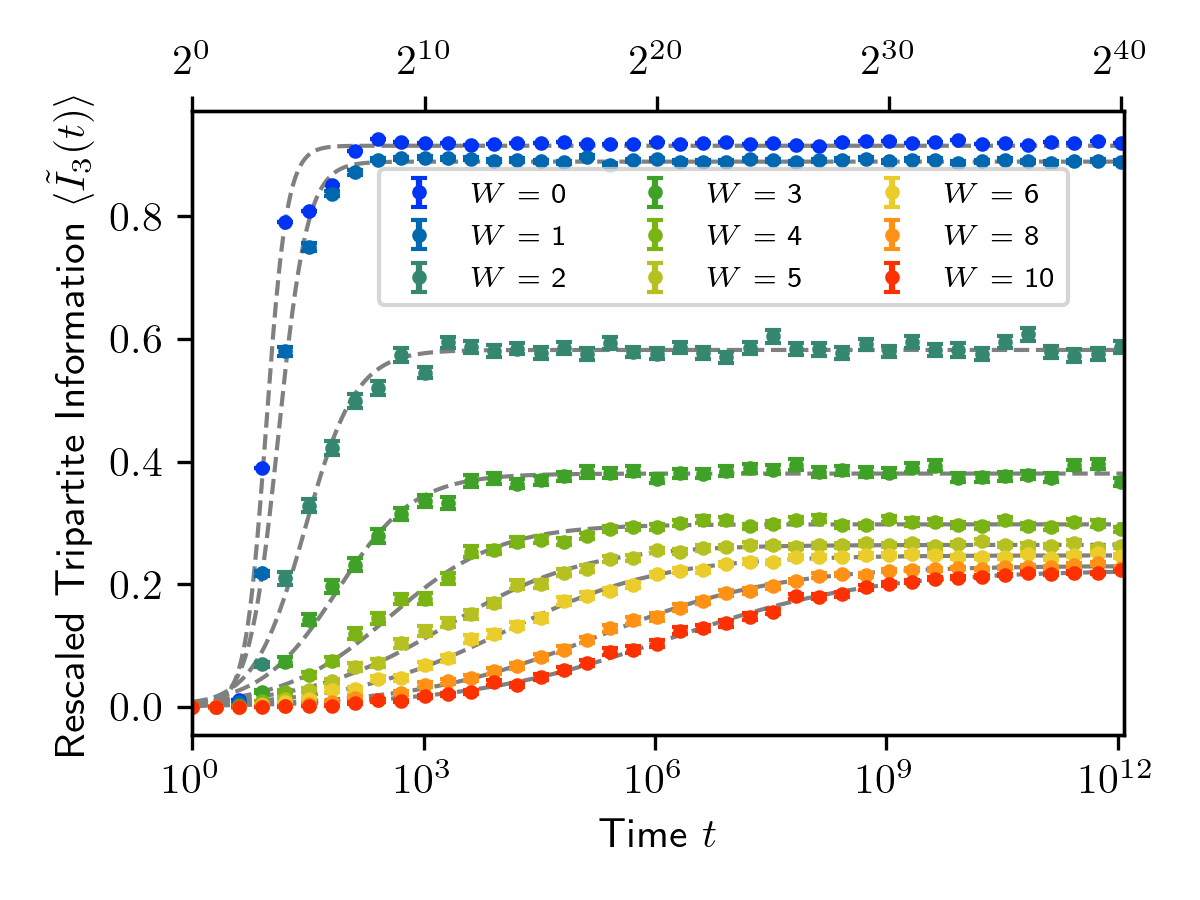}}
    \caption{Spreading of information through the isotropic Heisenberg chain ($\Delta = 1, L = 12$) between two subsystems at opposite ends. The values in Fig.\,\ref{fig:a1-logtime} were normalized because the step becomes very small for higher disorder strength $W$ and is hard to see otherwise. }
    \label{fig:a1-a7-logtime}
\end{figure}

Analogously to Fig.\,\ref{fig:a3-logtime} in the main part the rise of $I_3$ over a logarithmic timescale the behaviour is shown in Fig.\,\ref{fig:a1-a7-logtime} for $\ell=1$ and $\ell=3$. The shape of the transition from the initial to the asymptotic value is qualitatively the same in all three cases.

\begin{figure}[htb]
    \subcaptionbox{$\ell = 1$\label{fig:a1-asymptotic-change}}{\includegraphics[width=0.5\columnwidth]{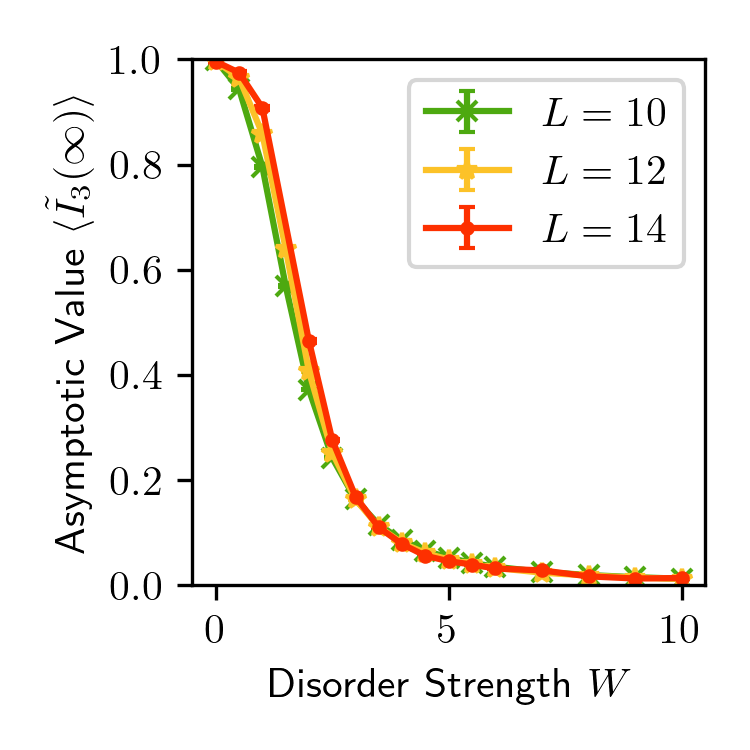}}\subcaptionbox{$\ell = 3$\label{fig:a7-asymptotic-change}}{\includegraphics[width=0.5\columnwidth]{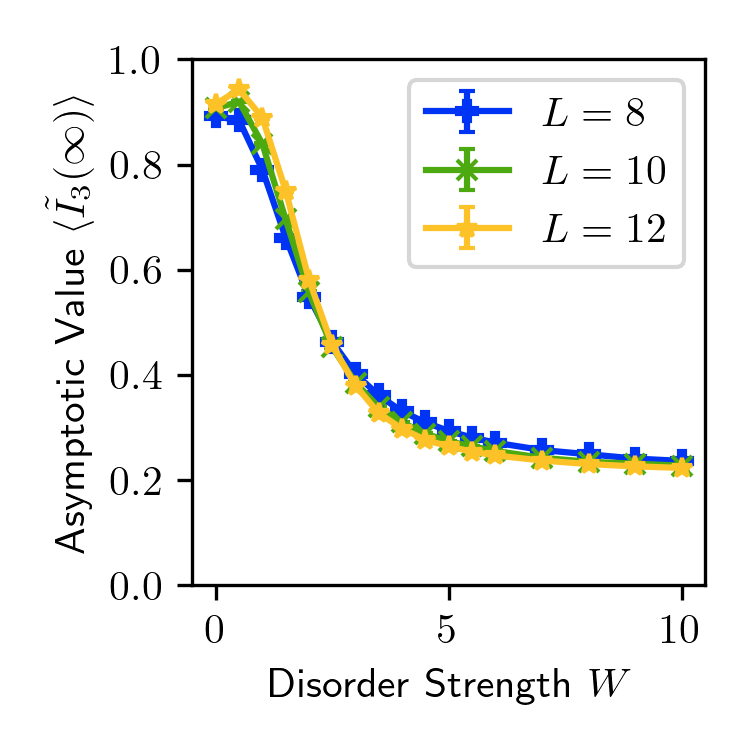}}
    \caption{Asymptotic value of tripartite information between two subsystems at opposite ends of the isotropic Heisenberg chain ($\Delta = 1$).}
    \label{fig:a1-a7-asymptotic-change}
\end{figure}

The value of the asymptotic value both in the clean system limit $W \to 0$ and in the high disorder limit depends on the subsystem size $\ell$ however, as can be seen in  Fig.\,\ref{fig:a1-a7-asymptotic-change} in analogy to Fig.\,\ref{fig:a3-asymptotic-change} in the main part.

\begin{figure}[htb]
    \subcaptionbox{$\ell = 1$\label{fig:a1-arrival-time-over-d}}{\includegraphics[width=0.5\columnwidth]{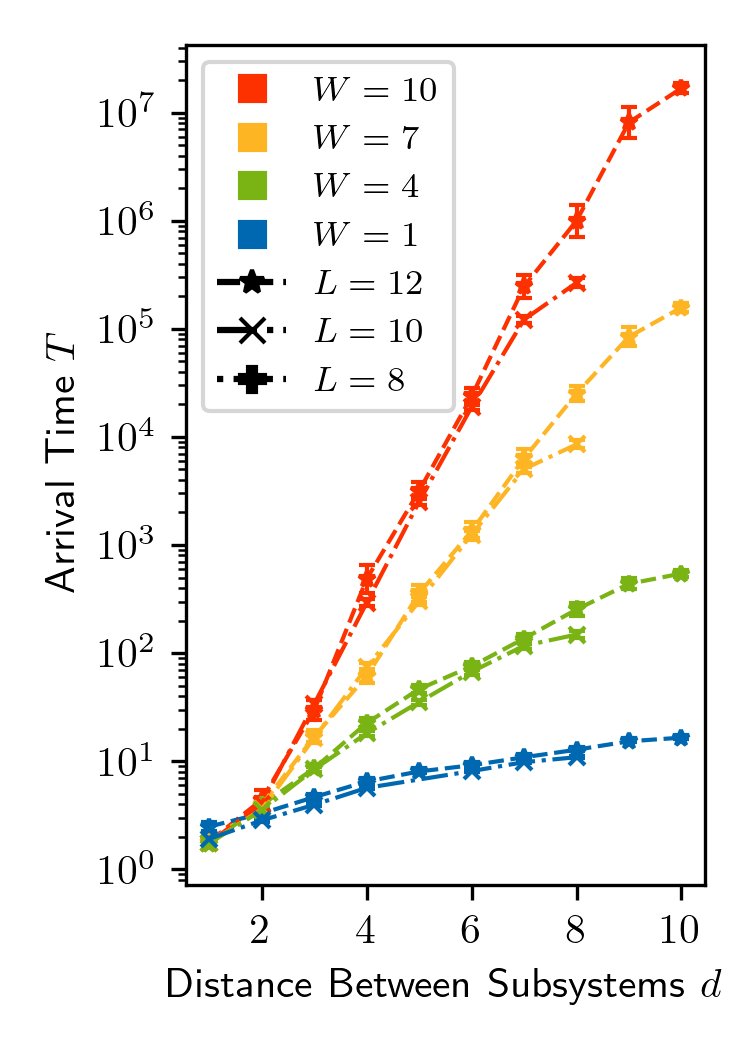}}\subcaptionbox{$\ell = 3$\label{fig:a7-arrival-time-over-d}}{\includegraphics[width=0.5\columnwidth]{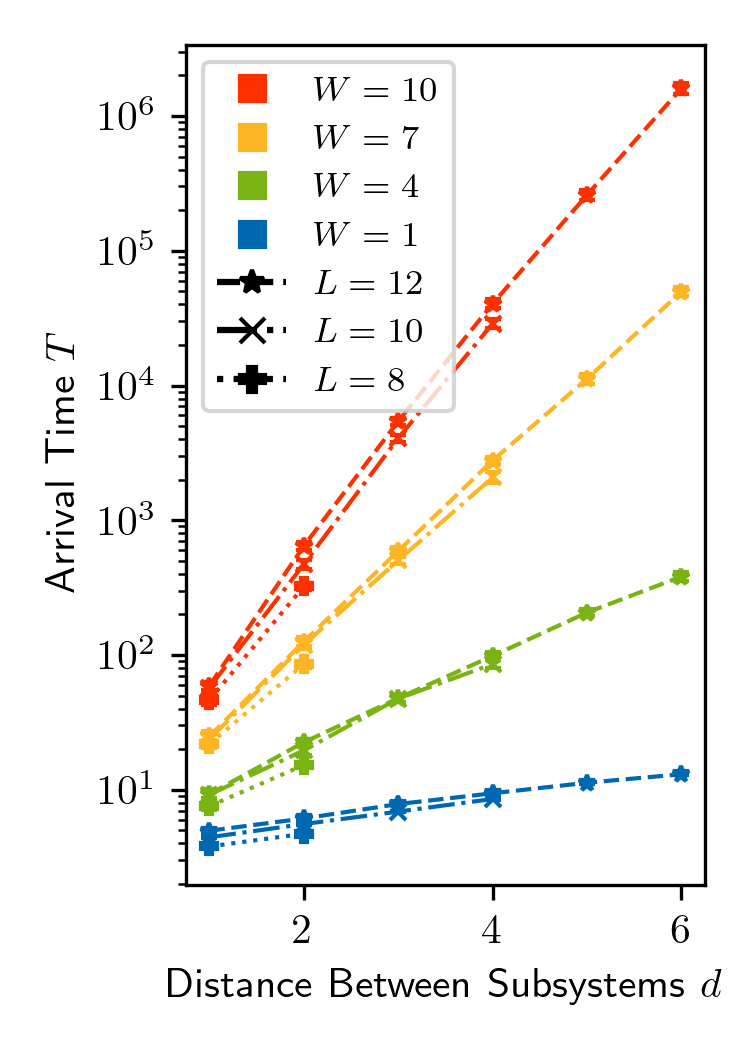}}
    \caption{Arrival time of the information between two subsystems of the isotropic Heisenberg chain ($\Delta = 1$) across a distance of $d$ spins.}
    \label{fig:a1-a7-arrival-time-over-d}
\end{figure}
 While the asymptotic value transitions from 1 to 0 for the single spin subsystem $\ell = 1$, the step becomes smaller for larger subsystem sizes $\ell$. As already discussed in the main part (cf. Fig.\,\ref{fig:asymptotic-finite-size}) we think the deviation from the Haar limit $\tilde I_3 = 1$ for low disorder is a finite size effect that should vanish in the $\ell / L \to 0$ limit. The dependence on the subsystem size $\ell$ for large disorder however is not clear yet, numerically it is caused by a rise in $I(\mathcal{A}:\mathcal{D})$ over time in Eq.\,\ref{eq:tripartite-information}.

The behaviour of the arrival time, as shown in Fig.\,\ref{fig:a1-a7-arrival-time-over-d} in analogy to Fig.\,\ref{fig:a3-arrival-time-over-d} in the main part is very similar in all cases, as the information travels inside a light cone that is growing logarithmically with time. The boundary effects – which suppress the arrival time if the distance between the subsystems is maximized on the open chain – become smaller as the subsystem size is increased, which makes intuitive sense, as the sites in a larger subsystem will be further from the end of the chain on average.
\section{BKT Scaling}
\label{sec:bkt-scaling}
\begin{figure}[htb]
    \includegraphics[width=\columnwidth]{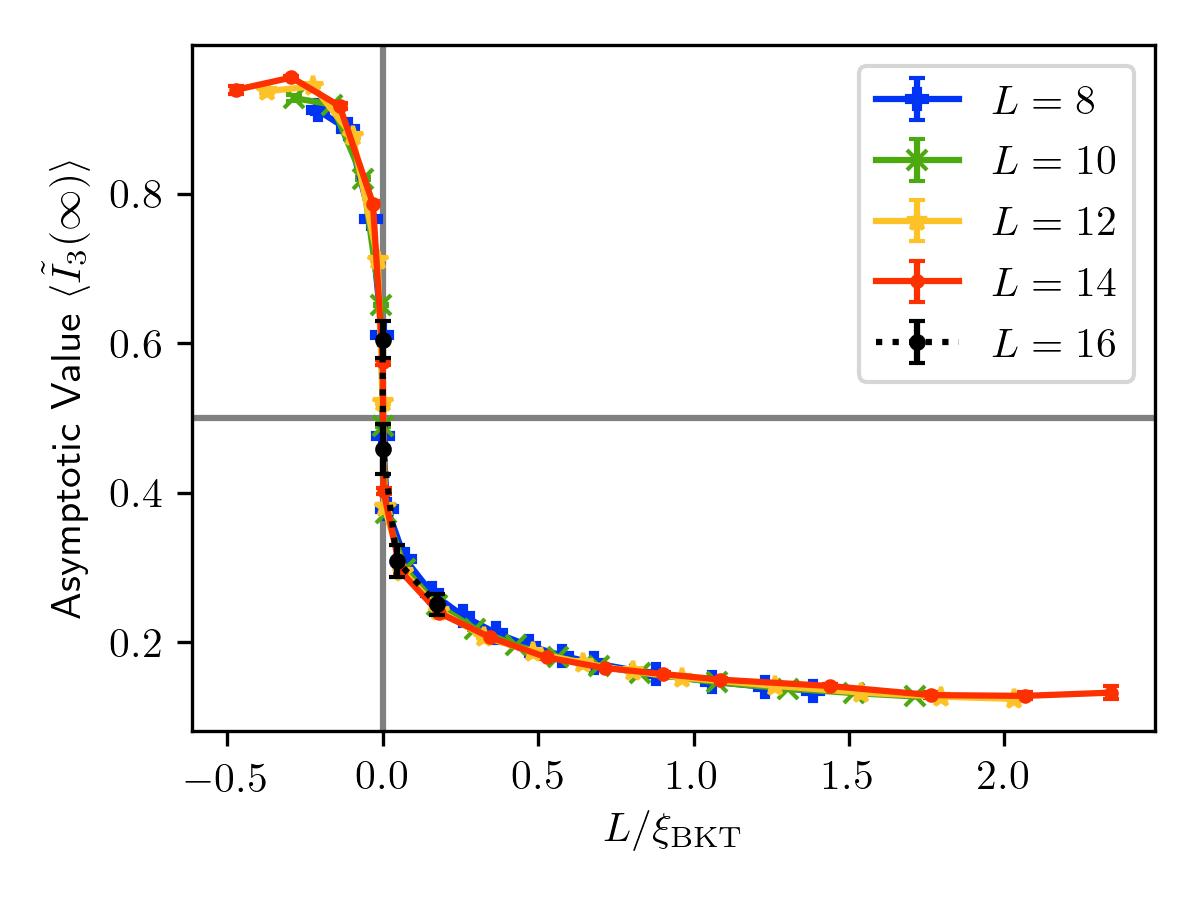}
    \caption{Asymptotic value at late times of the rescaled tripartite information $\tilde I_3(t)$ between two subsystems of size $\ell = 2$ at opposite ends of the isotropic Heisenberg chain ($\Delta = 1$).}
    \label{fig:a3-asymptotic-change-BKT}
\end{figure}

Following \citet{suntajsErgodicityBreakingTransition2020} we also fitted the asymptotic value of the tripartite information to a BKT-type curve in Fig.\,\ref{fig:a3-asymptotic-change-BKT}:
\begin{equation*}
\zeta_\mathrm{BKT} = \exp(b_\pm / \sqrt{|W-W*}|).
\end{equation*}

We found $b_+=b_- = 5.0$ to be a good match, with a fitted linear dependence of the critical disorder strength on the system size as follows:
\begin{equation*}
W^* = (3.25 \pm 0.12) + L (-0.02 \pm 0.02).
\end{equation*}
\end{appendix}
\end{document}